%% file: ms.tex
\documentclass[11pt,a4paper]{article}
\usepackage{jcappub}

\usepackage{amsmath,amssymb,bm,float,Macro,mathrsfs,tabularx}
\usepackage{color}
\usepackage{hyperref}

\let\normalcolor\relax
\input{local-macro}

\title{
	Lensing reconstruction from a patchwork of polarization maps 
} 

\author[a]{Toshiya Namikawa}
\author[b]{and Ryo Nagata}

\affiliation[a]{Yukawa Institute for Theoretical Physics, Kyoto University, Kyoto 606-8502, Japan}
\affiliation[b]{High Energy Accelerator Research Organization (KEK), Tsukuba, Ibaraki 305-0801, Japan}

\emailAdd{namikawa@yukawa.kyoto-u.ac.jp}
\emailAdd{rnagata@post.kek.jp}

\abstract{
The lensing signals involved in CMB polarization maps have already been measured with 
ground-based experiments such as SPTpol and POLARBEAR, and would become important as 
a probe of cosmological and astrophysical issues in the near future. 
Sizes of polarization maps from ground-based experiments are, however, limited by contamination 
of long wavelength modes of observational noise. 
To further extract the lensing signals, 
we explore feasibility of measuring lensing signals from a collection of small sky maps 
each of which is observed separately by a ground-based large telescope, i.e., 
lensing reconstruction from a patchwork map of large sky coverage organized from small sky patches. 
We show that, although the B-mode power spectrum obtained from the patchwork map is biased 
due to baseline uncertainty, bias on the lensing potential would be negligible 
if the B-mode on scales larger than the blowup scale of $1/f$ noise is removed 
in the lensing reconstruction. 
As examples of cosmological applications, we also show 1) the cross-correlations between 
the reconstructed lensing potential and full-sky temperature/polarization maps 
from satellite missions such as PLANCK and LiteBIRD, 
and 2) the use of the reconstructed potential for delensing B-mode polarization 
of LiteBIRD observation. 
}

\begin{document} 

\maketitle


\input{sec1}

\input{sec2}
\input{sec3}

\input{sec4}
\input{sec5}


\acknowledgments 
We thank Yuji Chinone for helpful comments on analysis of CMB polarization map. 
TN thanks Duncan Hanson and Ryan Keisler for discussion on delensing. 
This work was supported in part by JSPS Grant-in-Aid for Research Activity Start-up (No. 80708511). 
We acknowledge the use of 
{\tt Healpix} \citep{Gorski:2004by}, 
{\tt Lenspix} \citep{Challinor:2005jy} and 
{\tt CAMB} \citep{Lewis:1999bs}. 

\appendix
\input{appA}

\bibliographystyle{JHEP}
\bibliography{cite}

\end{document}

%% file: local-macro.tex
\def\grad{\phi}
\def\curl{\varpi}

\def\tB{\widetilde{B}}

\def\hE{\widehat{E}}
\def\hB{\widehat{B}}
\def\hX{\widehat{X}}
\def\hY{\widehat{Y}}

\def\tC{\widetilde{C}}
\def\hC{\widehat{C}}
\def\hCEE{\widehat{C}^{\rm EE}}

\def\hCBB{\widehat{C}^{\rm BB}}
\def\hCXX{\widehat{C}^{\rm XX}}

\def\hCYY{\widehat{C}^{\rm YY}}

\def\tCEE{\widetilde{C}^{\rm EE}}
\def\tCBB{\widetilde{C}^{\rm BB}}

\def\CEE{C^{\rm EE}}
\def\CBB{C^{\rm BB}}

\def\estg{\widehat{\grad}}

\def\bn{\bm{\nabla}}
\def\hatn{\hat{\bm{n}}}

%% file: sec1.tex
\section{Introduction}
\label{sec.1}

The map of cosmic microwave background (CMB) polarization produced from primordial density 
fluctuations at the cosmic recombination epoch has specific spatial pattern of even parity 
which is called as E-mode. 
Mass distribution between the last scattering surface and the Earth bends the path of CMB photons and 
disturbs the spatial pattern of the polarization map, which violates parity symmetry and induces odd 
parity pattern (so-called B-mode) (e.g., \cite{Zaldarriaga:1998ar}). 
Since the gravitational lensing distortion is non-linear effect in terms of perturbation, 
the lensed polarization map has off-diagonal correlations in angular multipole space which can 
be utilized for reconstruction of the gravitational lensing potential. 
The CMB polarization provides a powerful probe of the lensing mass distribution. 
The reconstructed lensing potential is a tracer of the evolution of large scale structure via the 
correlation with late-time integrated Sachs-Wolfe (ISW) effect in the CMB temperature fluctuations 
(e.g., \cite{Goldberg:1999xm,Seljak:1998nu}). 
One can further apply the reconstructed lensing potential to estimate the lensing B-mode 
and remove it \cite{Knox:2002pe,Kesden:2002ku,Seljak:2003pn,Smith:2010gu,Teng:2011xc} for improving 
the signal-to-noise of primordial gravitational waves. 

Recently, the mapping of the lensing mass distribution became realized by some CMB observations. 
PLANCK team performed lensing reconstruction from the full-sky CMB temperature map \cite{Ade:2013tyw}. 
On the other hand, the lensing potential is reconstructed from ground-based experiments such as 
SPTpol and POLARBEAR using a finite size of their polarization map \cite{Hanson:2013daa,PB1:2013a} 
(see e.g. \cite{Namikawa:2014xga} for recent progress in CMB lensing). 
Next generation projects of CMB observation such as 
CMBPol \footnote{\url{http://cmbpol.uchicago.edu/}}, 
COrE \footnote{\url{http://www.core-mission.org/}}, and 
PRISM \footnote{\url{http://www.prism-mission.org/}}, 
are planning to realize high-sensitivity measurement of CMB polarization to observe 
the B-mode polarization down to arcminute scales. 
Their target sensitivities are enough to measure the B-mode signal of small angular scales 
caused by gravitational lensing distortion and greatly improve the efficiency of lensing 
reconstruction. 

Efficient reconstruction of the lensing potential requires knowledge of small scale structure of 
arcminute scales. 
Therefore, a simple solution for comprehensive lensing reconstruction is to observe 
a large part of the whole sky, at the same time, with high angular resolution. 
Although such full-sky observation of the polarization map can be achieved by a large-size 
satellite mission, usually it requires a huge budget. 
Instead of such observation, we propose another way of an efficient lensing reconstruction; 
the lensing potential is reconstructed from a collection of small sky maps each of which is 
observed separately by a ground-based large telescope, namely lensing reconstruction 
from a patchwork map of large sky coverage organized from small sky patches. 
Although each partial sky map observed separately keeps its coherence only within 
the respective sky patch, what is crucial for lensing reconstruction is sensitivity to small scale 
structure in the map. 
It may be a feasible task to make a lensing potential map by ground-based observations 
which require a much smaller budget in total compared with a single large-size satellite observation. 
If the lensing potential reconstructed by the patchwork scheme still keeps the coherence on the 
scales larger than the sizes of constituent patches (usually several degrees), 
the cross-correlation with the CMB temperature fluctuations allows a way to investigate 
late-time ISW effect which has substantial contribution to the temperature anisotropy on 
the scales of several dozen degrees. 
Furthermore, if the reconstructed potential can be applied for evaluation of the lensing B-modes 
whose coherence scales are larger than the sizes of constituent patches, it is possible to 
provide a template for delensing of the large scale lensing B-modes. 

The purpose of this paper is to perform a simulation of the lensing reconstruction 
from a full-sky  patchwork polarization map and investigate the cosmological applicability of 
the reconstructed potential, in particular, to measurement of late-time ISW effect and also 
delensing of the large scale lensing B-modes. 
In our simulation, the whole sky is divided into (at most) a few thousand partial sky patches 
each of which shares a same CMB realization but has a different noise realization. 
Since the patchwork map is supposed to be made by ground-based observations, 
signal baselines of the constituent patches are totally contaminated by long wavelength components of 
observational noise which mostly come from atmospheric sources and observational apparatus itself. 
The long wavelength domain of the noise spectrum is dominated by so-called $1/f$ noise. 
We focus on the influence of incoherence 
between neighboring patches which is induced by such baseline uncertainties and 
plays a key role in our simulation. 

Throughout this paper, we assume a flat $\Lambda$CDM model characterized by six parameters which are 
the baryon density ($\Omega\rom{b}h^2$), non-relativistic matter density ($\Omega\rom{m}h^2$), 
dark energy density ($\Omega_{\Lambda}$), scalar spectral index ($n\rom{s}$), 
scalar amplitude defined at $k=0.05$Mpc$^{-1}$ ($A\rom{s}$), and 
reionization optical depth ($\tau$). 
The cosmological parameters have the best-fit values of PLANCK 2013 results \cite{Ade:2013zuv}; 
$\Omega\rom{b}h^2=0.0220$, $\Omega\rom{m}h^2=0.1409$, $\Omega_{\Lambda}=0.6964$, 
$n\rom{s}=0.9675$, $A\rom{s}=2.215\times10^{-9}$, and $\tau=0.0949$. 
The beam size of the patchwork map is assumed to be $4$ arcminutes FWHM which is similar 
to the beam size of the POLARBEAR telescope because the POLARBEAR project has a future plan 
to extend the number of telescopes and perform a wide-field CMB lensing survey 
\footnote{\url{http://cosmology.ucsd.edu/simonsarray.html}}. 
This beam size is also similar to that of POLAR Array \cite{Teng:2011xc}
\footnote{\url{http://polar-array.stanford.edu/}}. 
We assume the one specific value of the beam size in our patchwork map simulation.  
Indeed, once a beam size smaller than $10$ arcminutes, which corresponds to the turnover scale 
of the lensing B-mode, is attained, sensitivity of such experiment to small scale structure 
depends rather on noise level. 
We repeat the whole analysis varying noise level of the patchwork map 
(see Sec. \ref{sec.3} and \ref{sec.4} for more details). 

This paper is organized as follows: 
In Sec.~\ref{sec.2}, the procedure of our map simulation is described. 
In Sec.~\ref{sec.3}, we discuss the statistical properties of the B-mode polarization power spectrum 
and reconstructed lensing potential based on our patchwork map simulation. 
In Sec.~\ref{sec.4}, we show the cross-correlation between the reconstructed lensing potential and 
the CMB anisotropies. 
Also, we discuss the feasibility of delensing by the reconstructed potential and the improvement of 
constraints on the tensor-to-scalar ratio and tensor spectral index. 
Finally, Sec.~\ref{sec.5} is devoted to some discussion and our conclusion.

%% file: sec2.tex
\section{Map simulation} \label{sec.2}

In our simulation, we prepare a patchwork map organized from small subpatches 
which are supposed to be observed separately by ground-based experiments (such as POLARBEAR). 
This is the map for lensing reconstruction. 
Although the patchwork map shares the CMB realization with the coherent fullsky map, 
its noise realization is generated through a more complicated procedure 
which includes a simulation of $1/f$ noise and subsequent baseline subtraction. 
In this section, we describe the procedure of our patchwork map simulation. 

In this paper, we consider a simple case where the total area of the patchwork map corresponds to 
the whole sky for the purpose of understanding how the patchwork scheme affects lensing reconstruction 
and delensing without confusing it with other biases from, for instance, multipole transformation 
in the presence of sky border (e.g., \cite{Namikawa:2013,Pearson:2014qna}). 
The HEALPix pixelization parameter ({\tt nside}) is set to be $2048$, which corresponds to 
the pixel size of $1.72$ arcmin, so that we can confirm the convergence of our calculation.

\subsection{Sky partition} 

For defining subpatches, we simply follow the HEALPix partitioning mechanism \cite{Gorski:2004by}. 
The patch size of POLARBEAR's deep survey is nearly $3$ degrees. 
Its extension may be realized by some modulator instrument up to about $15$ degrees. 
We try three cases in which the sizes of the respective subpatches are $3.66, 7.33,$ and 
$14.7$ degrees. The corresponding values of {\tt nside} are $16$, $8$, and $4$. 
Note that the subpatch sizes do not relate to noise levels in our analysis 
because we assume that the patchwork map covers the whole sky. 

\subsection{CMB polarization map} 

In one realization of the patchwork map, each constituent subpatch shares the same realization 
of CMB polarization which is also shared by the corresponding coherent fullsky map. 
This map contains the signals of primordial E-modes and lensing. 

We generate lensed CMB maps using {\tt Lenspix} \footnote{http://cosmologist.info/lenspix/}. 
The distortion effect of lensing on the polarization anisotropies at the last scattering surface 
(primary anisotropies) is expressed by a remapping. 
Denoting the primary polarization anisotropies at position $\hatn$ on the last 
scattering surface as $[Q\pm\iu U](\hatn)$, the lensed anisotropies in a direction $\hatn$, 
are given by (e.g., \cite{Zaldarriaga:1998ar}) 
\al{
	[Q\pm\iu U](\hatn) &= [Q\pm\iu U](\hatn + \bm{d}(\hatn)) 
	\notag \\ 
		&= [Q\pm\iu U](\hatn) + \bm{d}(\hatn)\cdot\bn [Q\pm\iu U] (\hatn) + \mC{O}(|\bm{d}|^2)
	\,. \label{Eq:remap}
}
The two-dimensional vector $\bm{d}$ is the deflection angle which is in general decomposed into two 
quantities by the parity symmetry (e.g., \cite{Hirata:2003ka,Cooray:2005hm,Namikawa:2011cs}): 
\al{
	\bm{d} = \bn\grad + (\star\bn)\curl 
	\,. \label{Eq:deflection}
}
Here the first and second terms are gradient and curl modes, respectively, and 
$\star$ denotes an operator which rotates the angle of two-dimensional vector counterclockwise by 
$90$-degree. In our simulation, we set $\curl=0$ since the curl mode is not generated by 
the gravitational potential at the linear order, but we will discuss reconstruction of the curl mode 
as a null test in Sec.~\ref{sec.4}. 
Instead of the spin-$2$ quantity, the polarization anisotropies are usually decomposed into 
the rotationally invariant combination, i.e., the E and B mode polarizations 
(e.g., \cite{Zaldarriaga:1998ar}). In harmonics space, the E and B-modes are 
defined with the spin-$2$ spherical harmonics ${}_{\pm 2}Y_{\ell m}$ \cite{Hu:2000ee}): 
\al{
	[E \pm \iu B ]_{\ell m} = -\Int{}{\hatn}{_} {}_{\pm 2} Y_{\ell m}^*(\hatn) [Q\pm \iu U](\hatn) 
	\,. 
}
Similarly, the harmonic coefficients of the scalar quantity $\grad$, the so-called lensing potential, 
is given by 
\al{
	\grad_{LM} = \Int{}{\hatn}{_} {}_{0} Y_{LM}^*(\hatn) \grad (\hatn) 
	\,, 
}
where ${}_{0}Y_{LM}$ is the spin-$0$ spherical harmonics. 
Note that, from Eq.~\eqref{Eq:remap}, 
the lensed E and B modes are then given by (e.g., \cite{Hu:2000ee}) 
\al{
	\tilde{E}_{\ell m} &= E_{\ell m} + \sum_{\ell'm'}\sum_{LM} 
		\Wjm{\ell}{\ell'}{L}{m}{m'}{M} \grad_{LM}^* 
		\{\mC{S}^{(+)}_{\ell\ell'L}E_{\ell'm'}^* + \iu\mC{S}^{(-)}_{\ell\ell'L}B_{\ell'm'}^*\} 
	\label{Eq:Lensed-E} \,,
	\\ 
	\tilde{B}_{\ell m} &= B_{\ell m} + \sum_{\ell'm'}\sum_{LM} 
		\Wjm{\ell}{\ell'}{L}{m}{m'}{M} \grad_{LM}^* 
		\{\mC{S}^{(+)}_{\ell\ell'L} B_{\ell'm'}^* - \iu\mC{S}^{(-)}_{\ell\ell'L} E_{\ell'm'}^*\} 
	\label{Eq:Lensed-B} \,, 
} 
where the quantities, $\mC{S}^{(\pm)}_{\ell\ell'L}$ is given by 
\al{ 
	\mC{S}^{(\pm)}_{\ell\ell'L} 
		&= \frac{1\pm (-1)^{\ell+\ell'+L}}{2} \sqrt{\frac{(2\ell+1)(2\ell'+1)(2L+1)}{16\pi}}
	\notag \\ 
		&\qquad \times 
			[-\ell(\ell+1)+\ell'(\ell'+1)+L(L+1)]\Wjm{\ell}{\ell'}{L}{2}{-2}{0} 
	\,. 
} 
To simulate the lensed CMB polarization map, we compute the {\it unlensed} 
angular auto-/cross-power spectra of 
the E-mode polarization and lensing potential with {\tt CAMB} \cite{Lewis:1999bs}. 

At this stage, the generated CMB polarization map is still coherent over the whole sky. 
As described in the next subsection, baseline uncertainty induces incoherence between 
neighboring patches, which significantly contaminates the CMB signal of angular scales 
larger than the sizes of subpatches. 
Even after baseline subtraction, polarization maps of neighboring patches exhibit mutual 
discrepancy on their border.

\subsection{Noise map} 

\subsubsection{Noise spectrum} 

Since the polarization map of each constituent subpatch is supposed to be observed separately, 
we generate an independent noise map for each subpatch. 
We assume that sky scanning is isotropic and residual noise map after data processing is described 
as a random Gaussian field. 

Before proceeding, let us introduce $1/f$ noise. 
The intensity of incident radiation into detectors is measured in terms of electrical signal. 
In the time-ordered data, CMB signal is tiny fluctuation on the signal baseline 
which consists of low frequency noises due to atmospheric disturbance and 
thermal fluctuation of observational apparatus. 
Identification of low frequency (i.e. large angular) component of CMB signal is restricted by 
such baseline uncertainties which are often called as $1/f$ noise. 
We incorporate the $1/f$ noise into our analysis as long wavelength noise 
which has a power spectrum of inverse powerlaw. 

The total noise power spectrum is defined as 
\al{
	\mC{N}_{\ell}
		&\equiv \left(\frac{\Delta\rom{P}}{T\rom{CMB}}\right)^2
		\exp\left[\frac{\ell(\ell+1)\theta^2}{8\ln 2}\right]
		\left[1+\left(\frac{\ell\rom{knee}}{\ell}\right)^{\alpha}\right]
	\,. \label{noise}
}
$T\rom{CMB}=2.7$K is the mean temperature of CMB. The quantity $\theta$ is beam size. 
$\Delta\rom{P}$ is noise level of polarization measurement. 
Here, $\ell$ is an integer between $1$ and $4096$. 
The $1/f$ noise is characterized by the parameters $\alpha$ and $\ell\rom{knee}$. 
The fiducial values for our assumed ground-based experiment are 
$\Delta\rom{P}=6\mu$K-arcmin and $\theta=4$ arcmin. 
We also investigate the cases of two other noise levels which roughly correspond to those of other 
ground-based experiments such as ACT, SPT and POLAR-Array. 
For the transition scale to the $1/f$ noise, we assume that $\ell\rom{knee}=100$, $50$, and $25$ 
in the cases of which the subpatch sizes are $3.66$, $7.33$, and $14.7$ degrees, respectively. 
The subpatch sizes are adjusted so that 
the $1/f$ noises do not significantly contaminate CMB signal of spatial scales smaller than 
the respective subpatches. 
Finally, the exponent $\alpha$ is chosen to be $1$, $2$, or $3$. 
Although we tried the three cases of $\alpha$, as shown in Sec.~\ref{sec.4}, 
the choice of $\alpha$ did not make significant difference in our result. 

The mutually independent $1/f$ noises induce incoherence between neighboring subpatches. 
This ``patchwork'' noise map is superposed onto the CMB map. 

\subsubsection{Baseline subtraction} 

While we incorporate blowup of $1/f$ noise around the knee scale into our analysis, 
polarization maps of subpatches practically have much larger ``offsets'' 
as seen from the fact that the noise power spectrum diverges at long wavelength limit 
($\ell \rightarrow 0$). 
In actual data processing, such large offsets are removed through a data analysis pipeline. 
We model the effect of the baseline subtraction procedure following the prescription below. 

For each subpatch, we remove the offset by computing the average signal (including noise) within 
the subpatch and subtracting it from the map. 
Given simulated maps $X^{\rm sim}$ described above, the baseline-subtracted map is evaluated 
by the equation as follows: 
\al{
	\hX (\hatn) = X^{\rm sim}(\hatn) - \sum_i W_i(\hatn) X_i 
		= X^{\rm CMB}(\hatn) + \sum_i W_i(\hatn) X_i^{\rm noise}(\hatn) - \sum_i W_i(\hatn) X_i
	\,, \label{Eq:X-bs}
}
where $X^{\rm CMB}$ is the lensed CMB anisotropies and $X_i^{\rm noise}$ is the noise field 
at $i$-th subpatch. $X_i$ is the baseline at $i$-th subpatch estimated as follows: 
\al{
	X_i = \frac{1}{A_i}\Int{2}{\hatn}{_} W_i(\hatn) X^{\rm sim}(\hatn)
		= \frac{1}{A_i}\Int{2}{\hatn}{_} W_i(\hatn) [X^{\rm CMB}(\hatn) + X^{\rm noise}(\hatn)]
	\,. 
}
The function $W_i$ is given by 
\al{
	W_i(\hatn) = \begin{cases} 
			1 & (\hatn \in \text{$i$-th subpatch}) \\
			0 & (\hatn \not\in \text{$i$-th subpatch} ) 
		\end{cases}
	\,, 
}
and the quantity $A_i$ denotes the area of each subpatch: 
\al{
	A_i = \Int{2}{\hatn}{_} W_i(\hatn)
	\,. 
}

In the absence of baseline subtraction, the angular power spectrum of the patchwork map 
exhibits large ringing. 
On the other hand, some parts of CMB signal, which are almost homogeneous within respective patches, 
are lost by this procedure, which affects the accuracy of lensing reconstruction. 

%% file: sec3.tex
\section{CMB statistics from patchwork map} \label{sec.3}

In this section, we show effects of the baseline uncertainty on the B-mode power spectrum and 
lensing observables.

\subsection{B-mode power spectrum} 

\begin{figure} 
\bc
\includegraphics[width=75mm,clip]{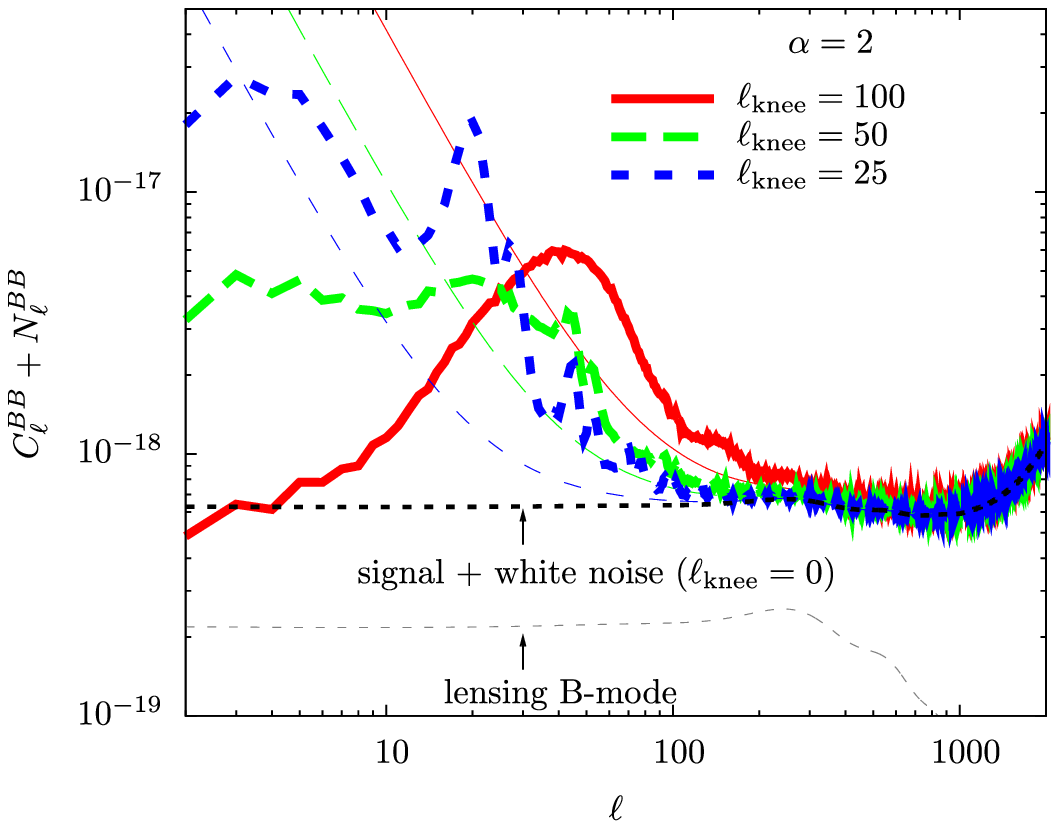}
\includegraphics[width=75mm,clip]{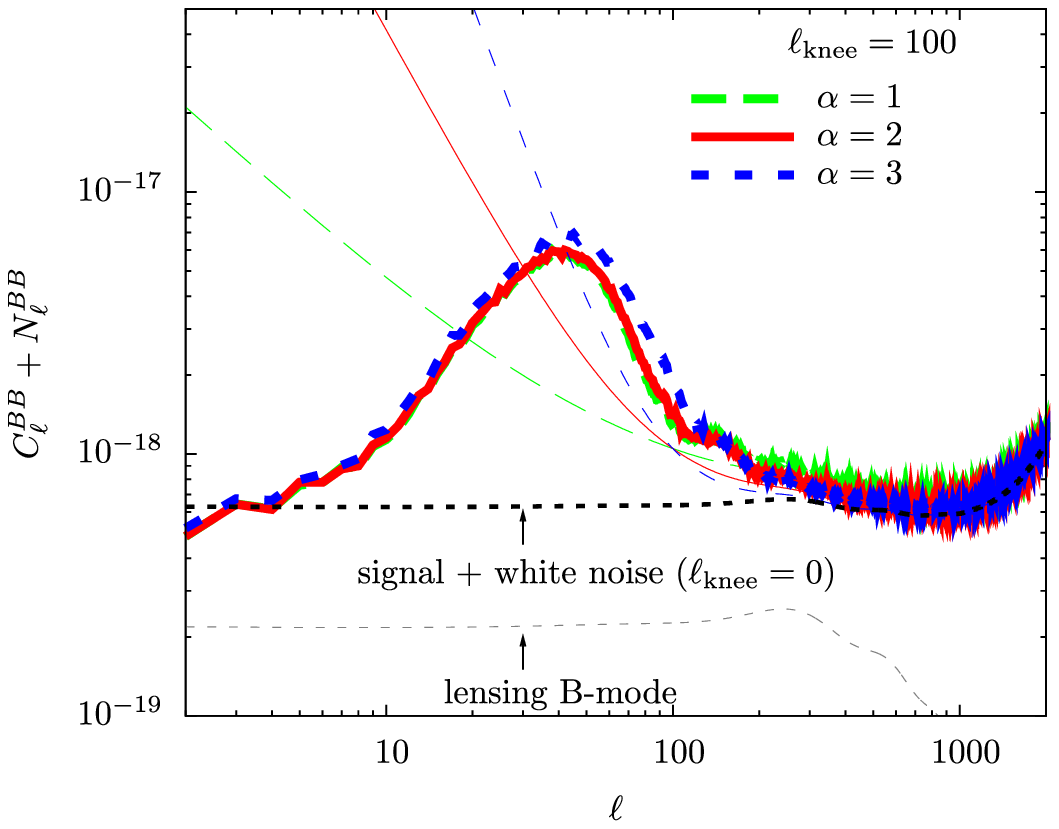}
\caption{
{\it Left}: B-mode angular power spectra after baseline subtraction (thick) compared with 
sums of the theoretical noise spectrum and lensing B-mode (thin), with varying 
$\ell\rom{knee}=100$, $50$, $25$ but fixing $\alpha=2$ (Left). 
The noise level ($\Delta\rom{P}$) is set to $6\mu$K-arcmin. 
We also show the case with the white noise (thick dashed) 
and the lensing B-mode power spectrum (thin dashed). 
{\it Right}: Same as the left panel but $\alpha$ is varied from $1$ to $3$ 
with $\ell\rom{knee}=100$. 
}
\label{Fig:Nls}
\ec
\end{figure}

\begin{figure} 
\bc
\includegraphics[width=75mm,clip]{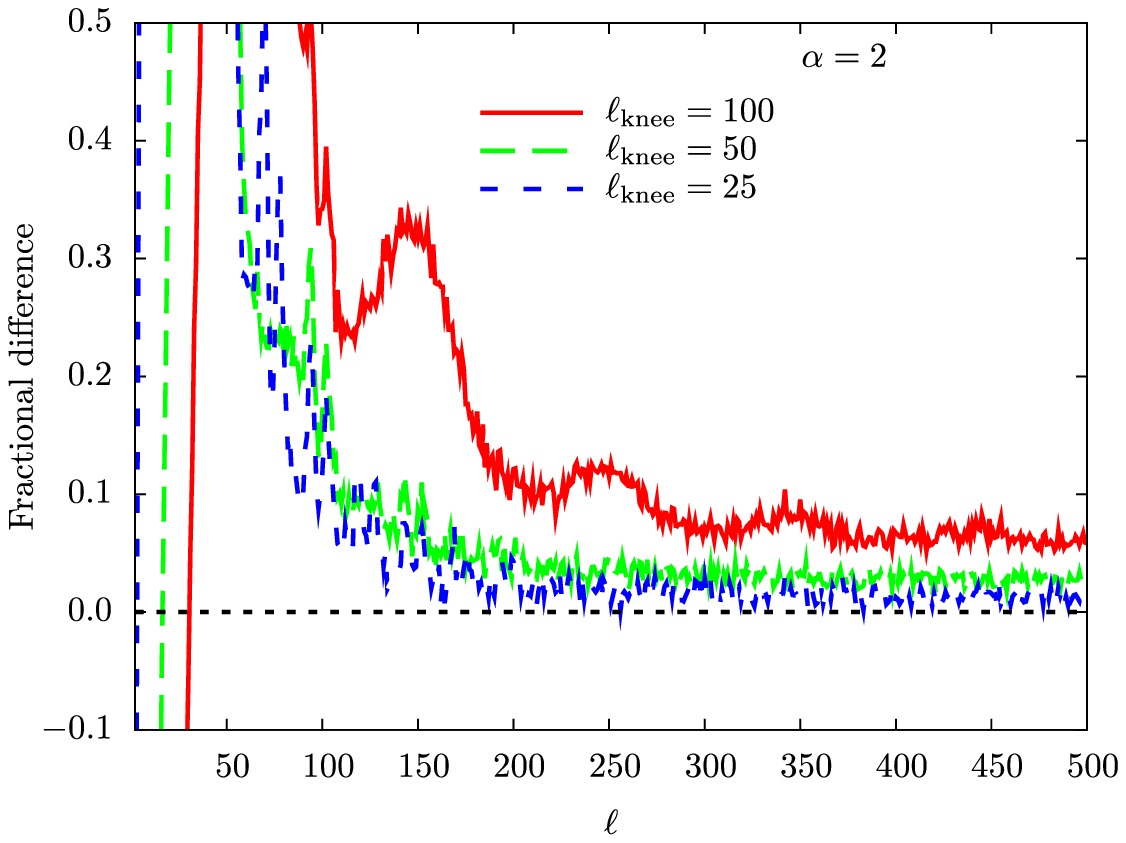}
\includegraphics[width=75mm,clip]{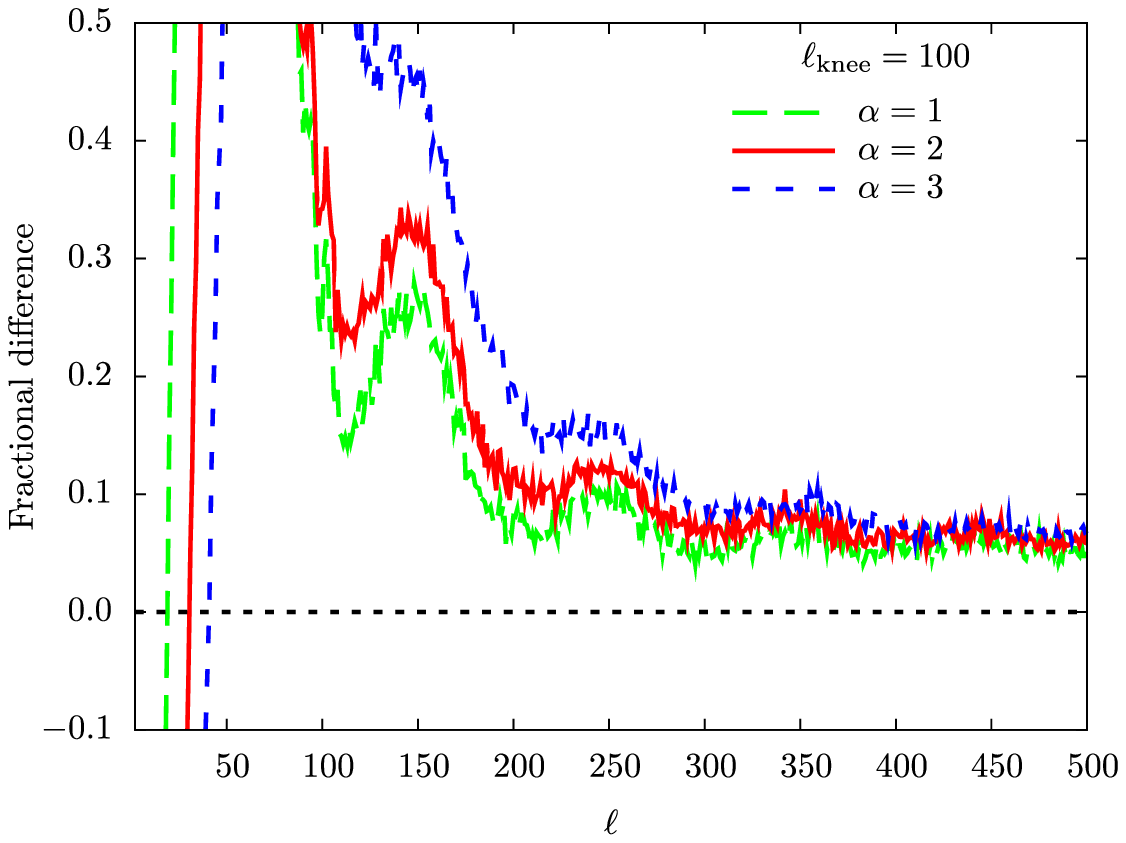}
\caption{
Same as Fig.~\ref{Fig:Nls} but for the fractional difference between simulated and theoretical 
B-mode power spectrum. 
}
\label{Fig:frac}
\ec
\end{figure}

\begin{figure} 
\bc
\includegraphics[width=75mm,clip]{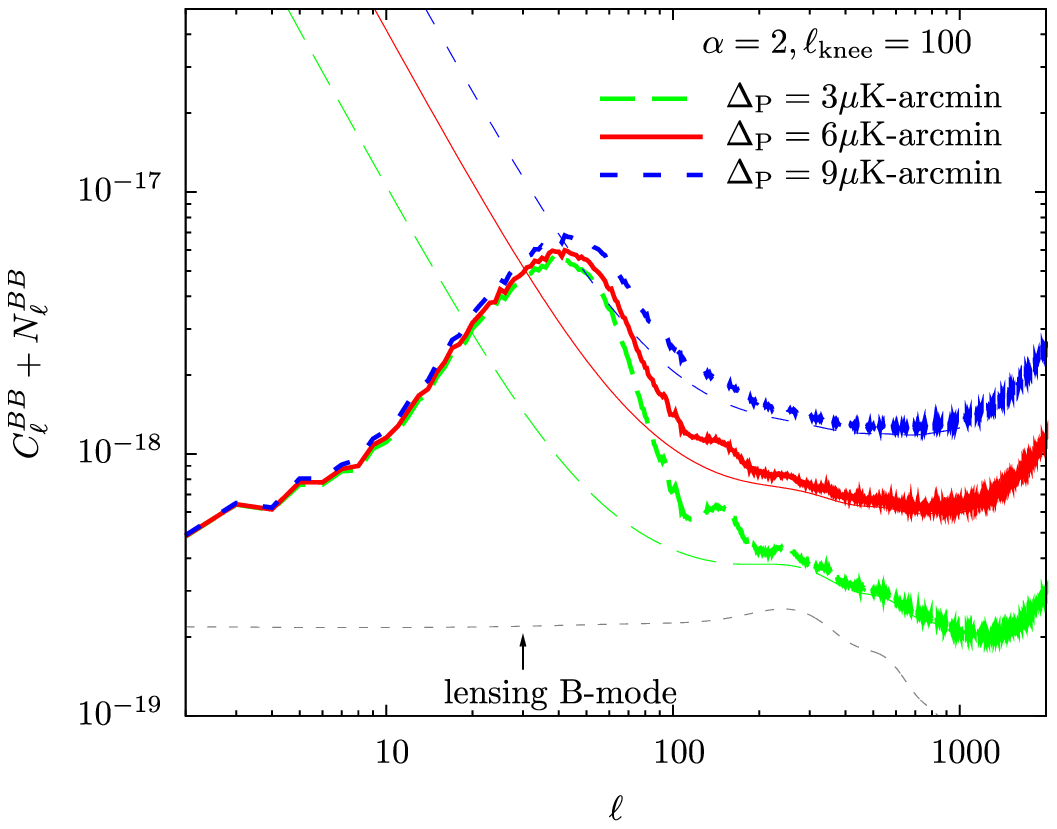}
\includegraphics[width=75mm,clip]{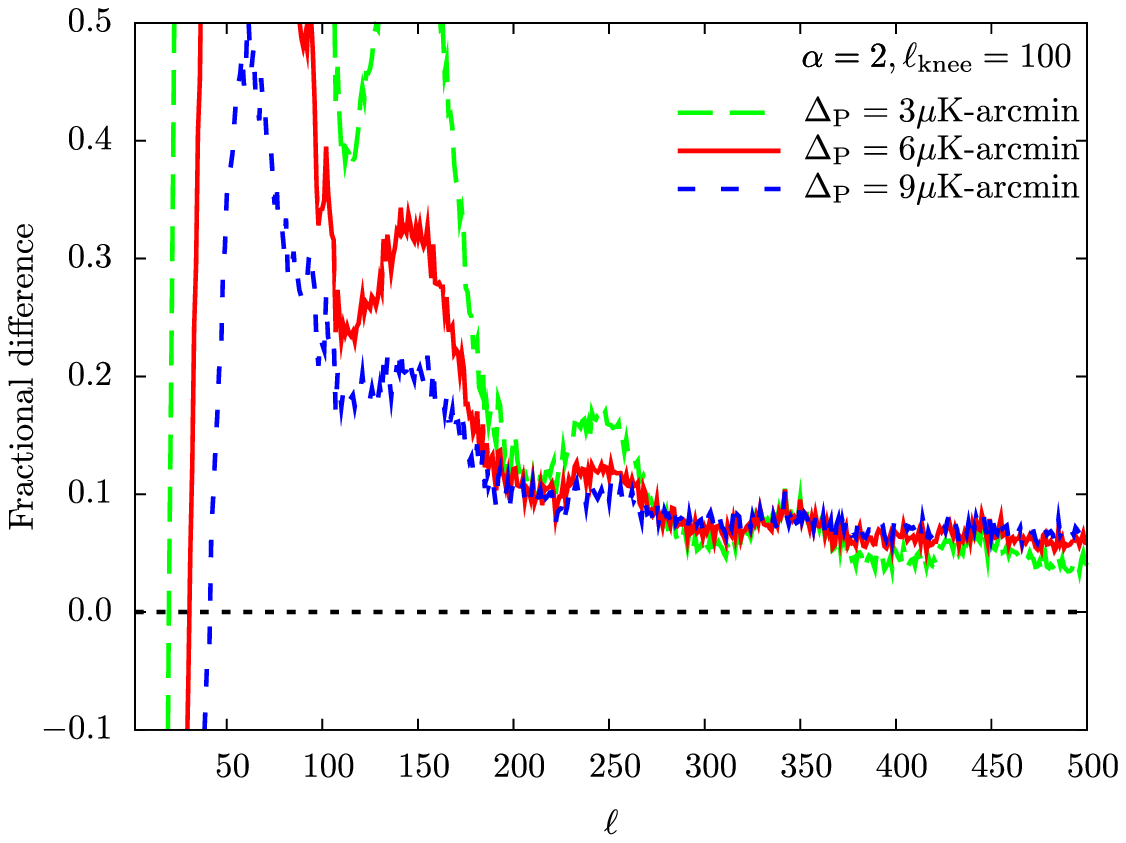}
\caption{
{\it Left}: Same as Fig.~\ref{Fig:Nls} but for different noise levels. 
{\it Right}: Same as Fig.~\ref{Fig:frac} but for different noise levels. 
}
\label{Fig:sigma}
\ec
\end{figure}

In Fig.~\ref{Fig:Nls}, we show B-mode power spectra obtained from the patchwork maps after 
baseline subtraction. The noise level ($\Delta\rom{P}$) is set to $6\mu$K-arcmin. 
In the left panel, the parameter $\ell\rom{knee}$ is varied as $100$, $50$ and $25$, 
while we fix $\alpha=2$. 
In the right panel, we fix $\ell\rom{knee}=100$ but $\alpha$ is varied as $1$, $2$ and $3$. 
Conventional spherical harmonic transformation results in significant bias on 
the B-mode power spectrum at large scales $\ell\leq\ell\rom{knee}$. 
The blowup scale of the bias depends on $\ell\rom{knee}$. On the other hand, the dependence on $\alpha$ is not so significant. 

In Fig.~\ref{Fig:frac}, we plot the fractional difference defined as $C^{\rm BB,sim}_{\ell}/C^{\rm BB,model}_{\ell}-1$ where $C^{\rm BB,sim}_{\ell}$ 
is obtained from the numerical simulations while $C^{\rm BB,model}_{\ell}$ is the sum of the theoretical noise spectrum \eqref{noise} and lensing B-mode. 
The fractional difference is significant at $\ell\lsim\ell\rom{knee}$, and is still $\sim 10\%$ on scales smaller than $\ell\rom{knee}$. 
In the case of Fig.~\ref{Fig:frac}, the lensing contributions involved in the B-mode power spectrum 
are $\sim 30\%$ at $\ell\gsim 100$ which is comparable or greater than the bias due to 
the baseline uncertainty. If the subpatch size becomes small, the fractional difference becomes large. 
These results imply that, even after the baseline subtraction, residual incoherence between 
subpatches still contributes to smaller scale modes through convolution with the window function $W_i$ which determines the subpatch size. 

In Fig.~\ref{Fig:sigma}, we also show the case with varying the noise level $\Delta\rom{P}$ from $9\mu$K-arcmin to $3\mu$K-arcmin. 
As the noise level $\Delta\rom{P}$ decreases, the bias around $\ell\sim\ell\rom{knee}$ increases. 
On the other hand, the bias at $\ell\gg\ell\rom{knee}$ is reduced by decreasing $\Delta\rom{P}$.

\subsection{Reconstructed deflection angle} 

\subsubsection{Quadratic lensing reconstruction} 

Next we consider the effect of baseline uncertainties on estimates of the deflection angle. 
Given lensed polarization anisotropies, $\tilde{X}_{\ell m}$ and $\tilde{Y}_{\ell m}$, 
the lensing effect induces the off-diagonal elements of the covariance 
($\ell\not=\ell'$ or $m\not=m'$): 
\al{
	\ave{\tilde{X}_{\ell m}\tilde{Y}_{\ell'm'}}\rom{CMB} 
		&= \sum_{L,M}\Wjm{\ell}{\ell'}{L}{m}{m'}{M}f^{\rm XY}_{\ell\ell'L}\grad^*_{LM} 
	\,, \label{Eq:weight} 
}
where $\ave{\cdots}\rom{CMB}$ denotes the ensemble average over the primary CMB anisotropies 
with a fixed realization of the lensing potential, and we ignore the higher-order terms 
of the lensing fields, $\mC{O}(\grad^2)$. 
For EE and EB, the weight function $f_{\ell\ell'L}^{\rm XY}$ in Eq.~\eqref{Eq:weight} 
is given by \cite{Okamoto:2003zw} 
\al{
	f^{\rm EE}_{\ell\ell'L} &= \mC{S}^{(+)}_{\ell\ell'L}\CEE_{\ell'} 
		+ \mC{S}^{(+)}_{\ell'\ell L}\CEE_{\ell} 
	\,, \\
	f^{\rm EB}_{\ell\ell'L} &= \iu [\mC{S}^{(-)}_{\ell\ell'L}\CBB_{\ell'} 
		+ \mC{S}^{(-)}_{\ell'\ell L}\CEE_{\ell}] 
	\,. 
}
Here $\CEE_{\ell}$ and $\CBB_{\ell}$ are the primary E and B-mode power spectrum, respectively. 
With a quadratic combination of observed polarization anisotropies, $\hX$ and $\hY$, 
Eq.~\eqref{Eq:weight} leads to the lensing estimators as (e.g., \cite{Okamoto:2003zw}), 
\al{
	[\estg^{\rm XY}_{LM}]^* 
		= A^{\rm XY}_L\sum_{\ell\ell'}\sum_{mm'}\Wjm{\ell}{\ell'}{L}{m}{m'}{M}
		g^{\rm XY}_{\ell\ell'L}\hX_{\ell m}\hY_{\ell'm'}
	\,. \label{Eq:estg-XY}
}
Here the quantity $g^{\rm XY}_{\ell\ell'L}$ and (diagonal) normalization $A^{\rm XY}_L$ 
are given by 
\al{
	g^{\rm XY}_{\ell\ell'L} 
		&= \frac{[f^{\rm XY}_{\ell\ell'L}]^*}{\Delta^{\rm XY}\hCXX_{\ell}\hCYY_{\ell'}} 
	\label{Eq:g-func} \\ 
	A^{\rm XY}_L 
		&= \left\{\frac{1}{2L+1}\sum_{\ell\ell'}
			f^{\rm XY}_{\ell\ell'L}g^{\rm XY}_{\ell\ell'L} \right\}^{-1}
	\,, \label{Eq:Rec:N0}
}
where $\Delta^{\rm EE}=2$, $\Delta^{\rm EB}=1$, and $\hCXX_{\ell}$ ($\hCYY_{\ell}$) is 
the observed power spectrum. 
The lensing reconstruction is then performed using the optimal 
combination of the EE and EB quadratic estimators: 
\al{
	\estg_{LM} = A_L\left(\frac{1}{A_L^{\rm EE}}\estg_{LM}^{\rm EE} 
		+ \frac{1}{A_L^{\rm EB}}\estg_{LM}^{\rm EB}\right)
	\,, \label{Eq:estg}
}
with $A_L^{-1} \equiv (A_L^{\rm EE})^{-1}+(A_L^{\rm EB})^{-1}$. 
Throughout this paper, to mitigate $\grad^4$-order bias \cite{Hanson:2010rp}, 
the lensed power spectrum ($\tCEE_{\ell}$ and $\tCBB_{\ell}$) is used in Eq.~\eqref{Eq:weight} 
rather than the primary one \cite{Lewis:2011fk,Anderes:2013jw}. 
For EB-quadratic estimator, we ignore the B-mode power spectrum in the weight function 
since it affects negligible contributions to the lensing estimator \cite{Okamoto:2003zw}. 
Note that other non-lensing anisotropies such as the residual contamination of baseline uncertainty 
at each subpatch can also generate the off-diagonal elements, and lead to bias on estimating 
the lensing potential as shown Sec.~\ref{sec:rec}. 

\subsubsection{Effect of baseline uncertainties on lensing observables} \label{sec:rec}

\begin{figure}
\bc
\includegraphics[width=75mm,clip]{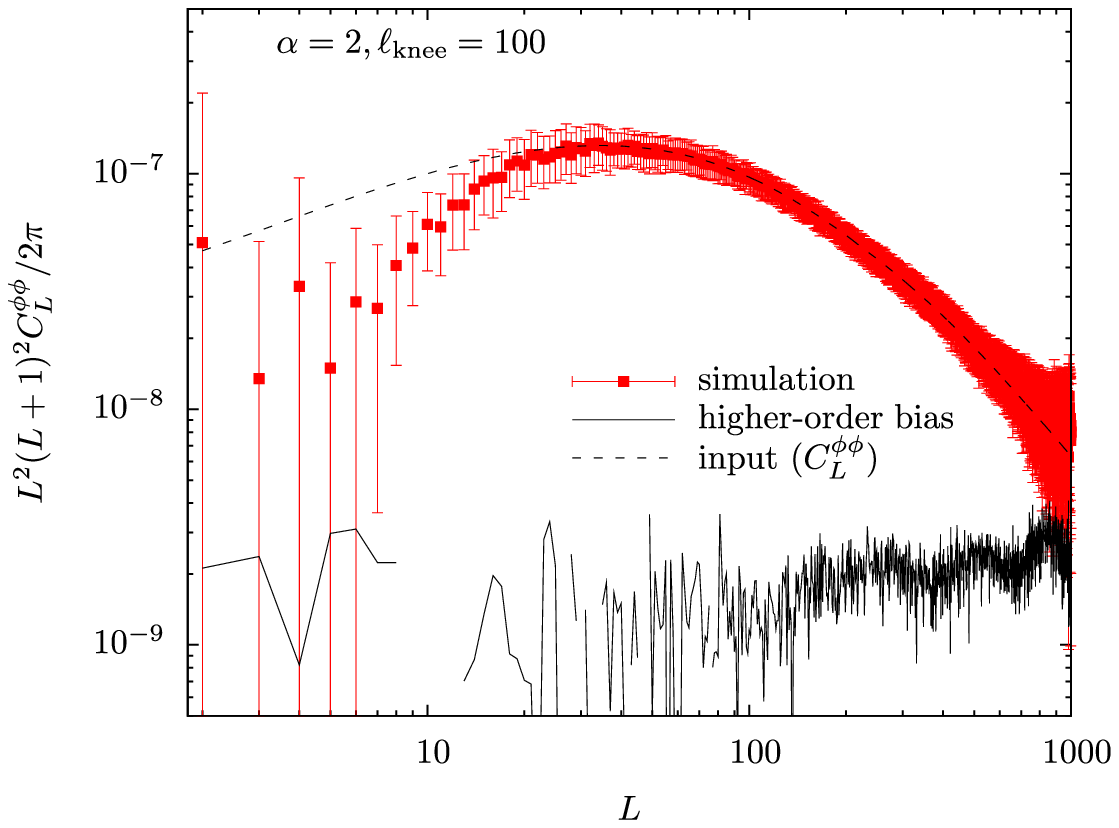}
\includegraphics[width=75mm,clip]{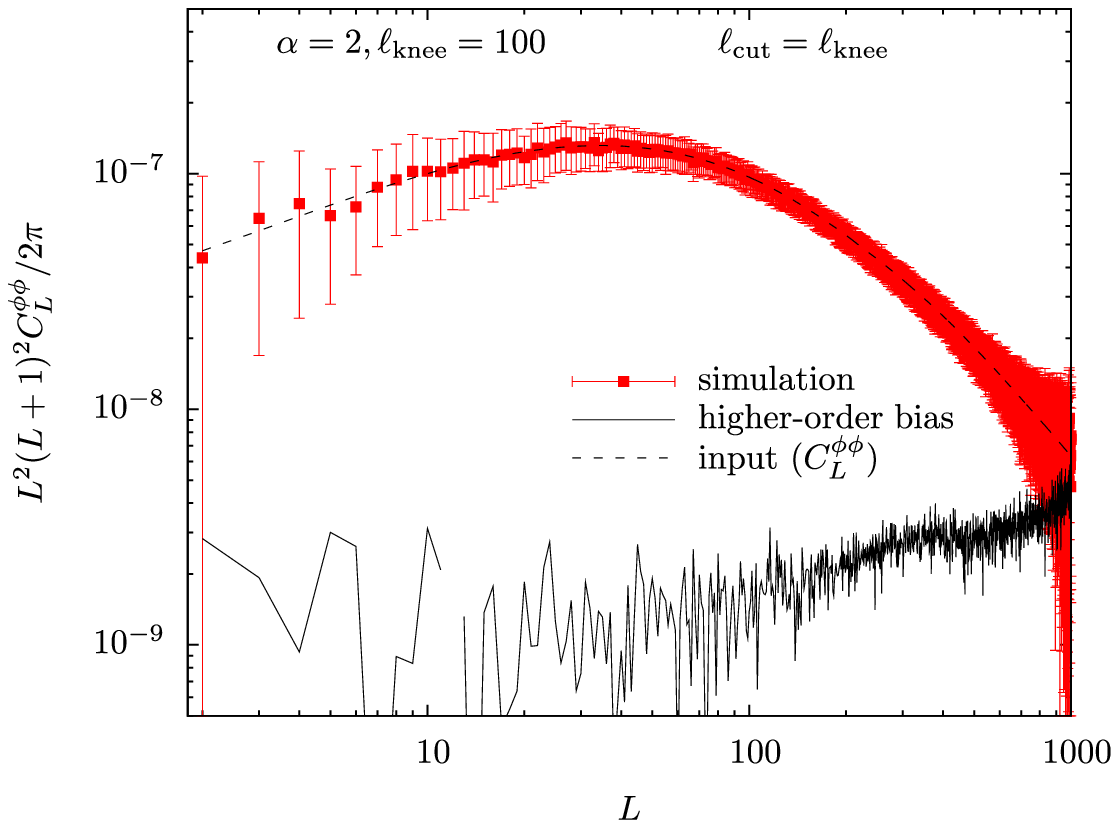}
\caption{
{\it Left}: 
Angular power spectrum of the lensing potential estimated from 
the patchwork maps with $\alpha=2$ and $\ell\rom{knee}=100$ (red points). 
The noise level ($\Delta\rom{P}$) is set to $6\mu$K-arcmin. 
The error bars show the variance of $C_{\ell}^{\grad\grad}$ evaluated from $100$ realizations of 
our simulated maps. 
In the lensing reconstruction, we use the E and B-mode multipoles at $2\leq \ell\leq 2000$. 
We also show the higher-order biases estimated from numerical simulations with $6\mu$K-arcmin and 
$4$ arcmin instrumental noise (solid). 
{\it Right}: 
Same as the left panel, but the E and B-modes at 
$\ell<\ell\rom{knee}$ are not used in the reconstruction. 
}
\label{Fig:rec-g}
\ec
\end{figure}

\begin{figure}
\bc
\includegraphics[width=75mm,clip]{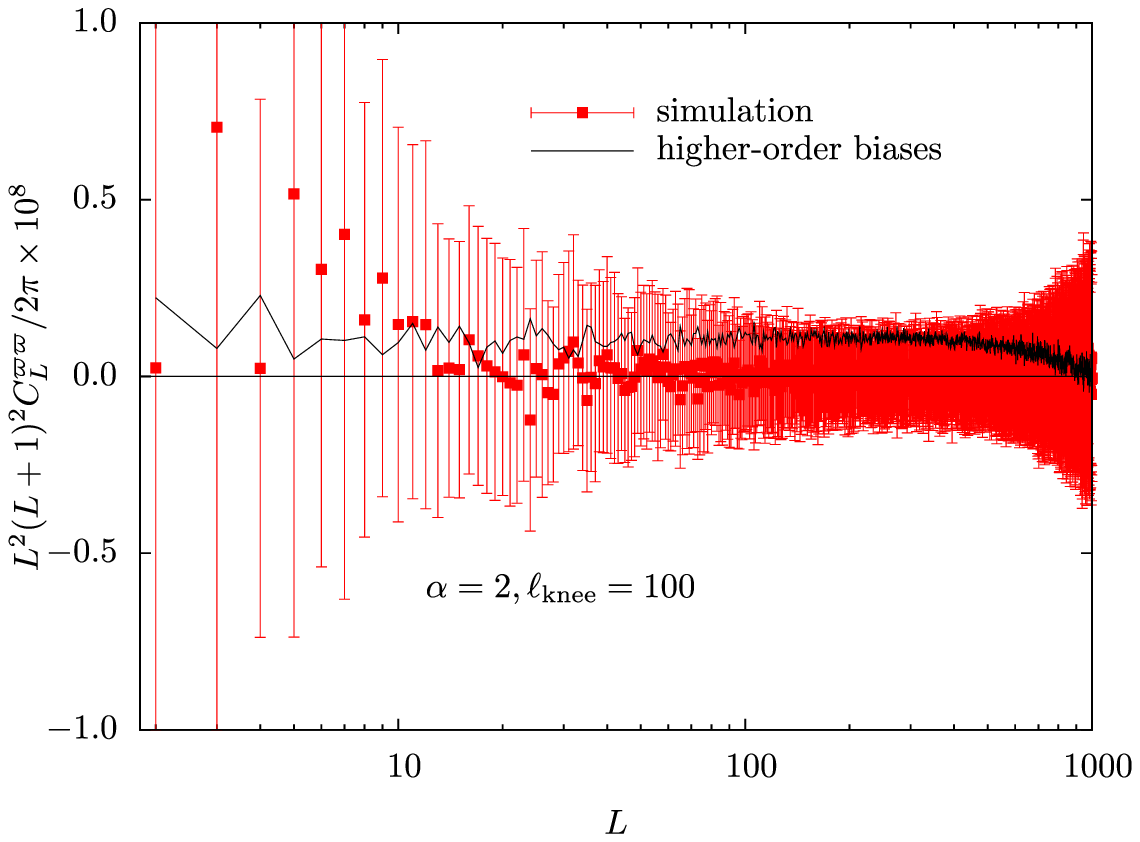}
\includegraphics[width=75mm,clip]{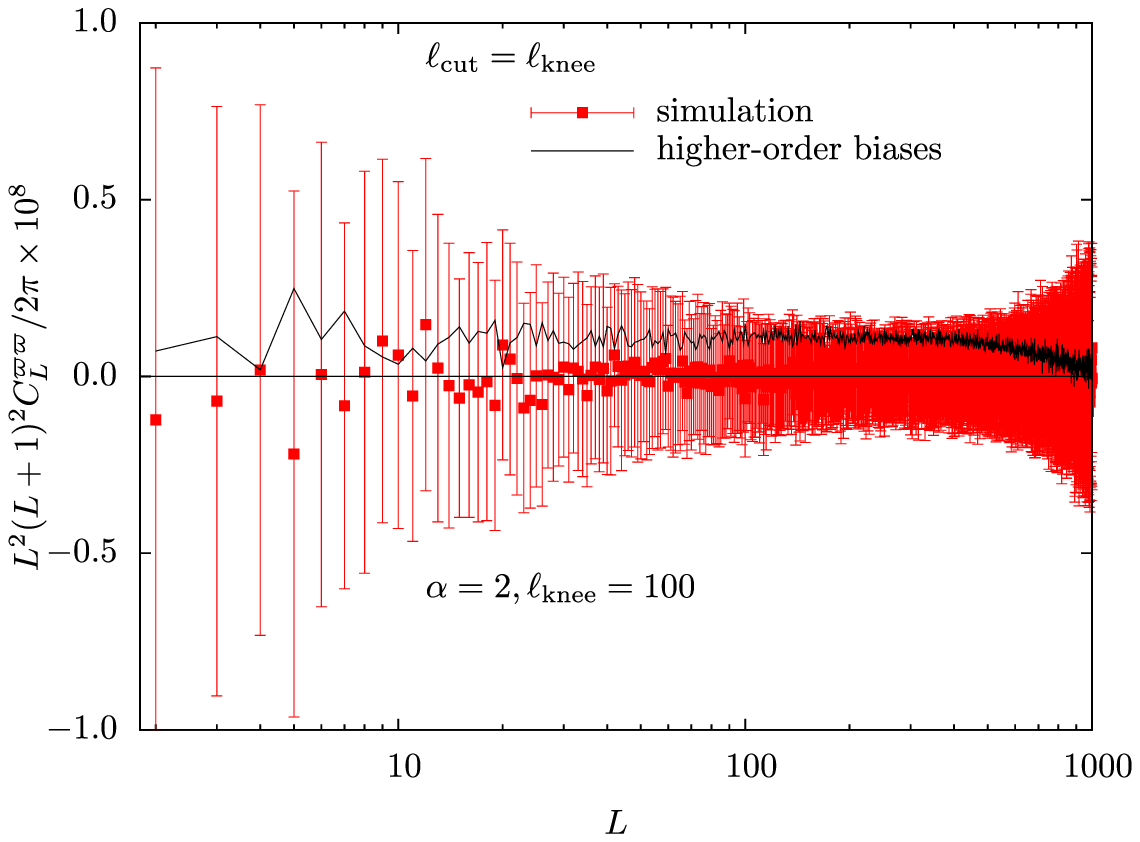}
\caption{
Same as Fig.~\ref{Fig:rec-g} but for curl mode. 
}
\label{Fig:rec-c}
\ec
\end{figure}

Let us discuss the effect of the baseline uncertainties on the estimation of the lensing power 
spectrum. From the lensing potential estimator \eqref{Eq:estg}, the lensing power spectrum 
is estimated through (see e.g. \cite{Ade:2013tyw,Namikawa:2014xga}) 
\al{
	\hC^{\grad\grad}_L = \frac{1}{2L+1}\sum_{M=-L}^{L}|\estg_{LM}|^2 - \widehat{N}^{(0)}_L - N_L^{(1)} 
	\,. 
}
Here the quantity $\widehat{N}_L^{(0)}$, the so-called Gaussian bias, denotes a correction for 
the disconnected part of the four-point correlations, and, in this paper, 
we simply estimate the Gaussian bias term as $\widehat{N}_L^{(0)}=A_L$ 
instead of the realization dependent approaches (\cite{Dvorkin:2008tf,Hanson:2010rp,Namikawa:2012pe,Namikawa:2013}). 
The quantity $N_L^{(1)}$ corrects for the bias terms arising from the secondary contractions of 
the lensing trispectrum \cite{Kesden:2003cc}, usually referred to as N1-bias. 
In this paper, we estimate the N1-bias from simulated maps in which the noise is assumed to have a white spectrum \cite{Anderes:2013jw}. 

In Fig.~\ref{Fig:rec-g}, we show the angular power spectrum of the lensing potential obtained from 
the patchwork maps in the case with $\alpha=2$ and $\ell\rom{knee}=100$. 
The noise level ($\Delta\rom{P}$) is set to $6\mu$K-arcmin. 
In the left panel, we perform the lensing reconstruction using CMB multipoles at $2\leq\ell\leq 2000$. 
If we naively use the B-mode at all scales in estimating 
the lensing potential ($2\leq \ell\leq 2000$), the power spectrum of the lensing estimator is biased on large scales ($L\,\lsim 20$). 
On the other hand, in the right panel, we show the same plot 
but the minimum multipoles of the E and B-modes used for the reconstruction $\ell\rom{cut}$ are set to $\ell\rom{knee}$. 
We find that the condition, $\ell\rom{cut}=\ell\rom{knee}$, would be enough to recover the lensing power spectrum. 
We also checked other cases of $\alpha$, $\ell\rom{knee}$ and $\Delta\rom{P}$, and find that 
$\ell\rom{cut}=\ell\rom{knee}$ is enough to reproduce the lensing power spectrum. 
This implies that, even if the B-mode power spectrum is biased due to the residual baseline uncertainty, the reconstructed lensing potential 
would be not so biased. This is because the bias on the B-mode power spectrum appeared in Fig.~\ref{Fig:frac} 
is mostly absorbed into the observed power spectrum involved in $g_{\ell\ell'L}^{\rm XY}$ (see \eqref{Eq:g-func}). 

Next we consider the curl mode of the deflection angle defined in Eq.~\eqref{Eq:deflection}. 
In the future CMB lensing analysis, not only the lensing potential (i.e. gradient mode) but also the curl mode would become important 
to probe cosmological sources of the non-scalar metric perturbations such as primordial gravitational waves 
and cosmic strings (see e.g. \cite{Namikawa:2013wda}), and/or as a cross check of the lensing potential reconstruction \cite{Ade:2013tyw}. 
In Fig.~\ref{Fig:rec-c}, to see whether the curl mode is consistent with zero, we show the reconstructed curl-mode power spectrum, 
in which we estimate the curl mode $\curl$ following full-sky formula of Ref.~\cite{Namikawa:2011cs}. 
Compared with the lensing potential, the curl mode is not so biased even if $\ell\rom{cut}=2$. 
Note that the power spectrum of the curl mode estimator also has the higher-order biases even in the absence of the curl mode sources, 
and the contributions of those terms in a temperature-based reconstruction are evaluated in Refs.~\cite{vanEngelen:2012va,BenoitLevy:2013bc}. 
As shown in Fig.~\ref{Fig:rec-c}, even in the polarization-based reconstruction, 
the higher-order biases on the curl mode would also have non-negligible contributions. 

\begin{figure}
\bc
\includegraphics[width=75mm,clip]{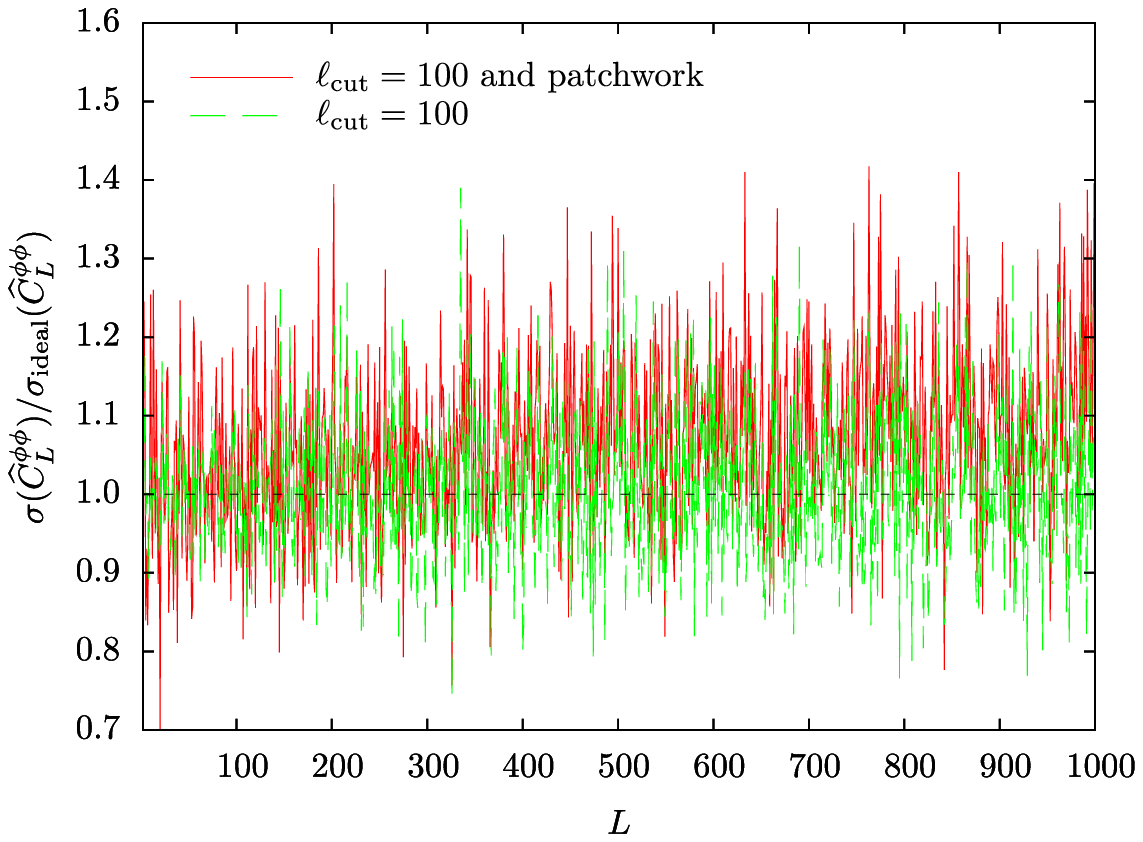}
\includegraphics[width=75mm,clip]{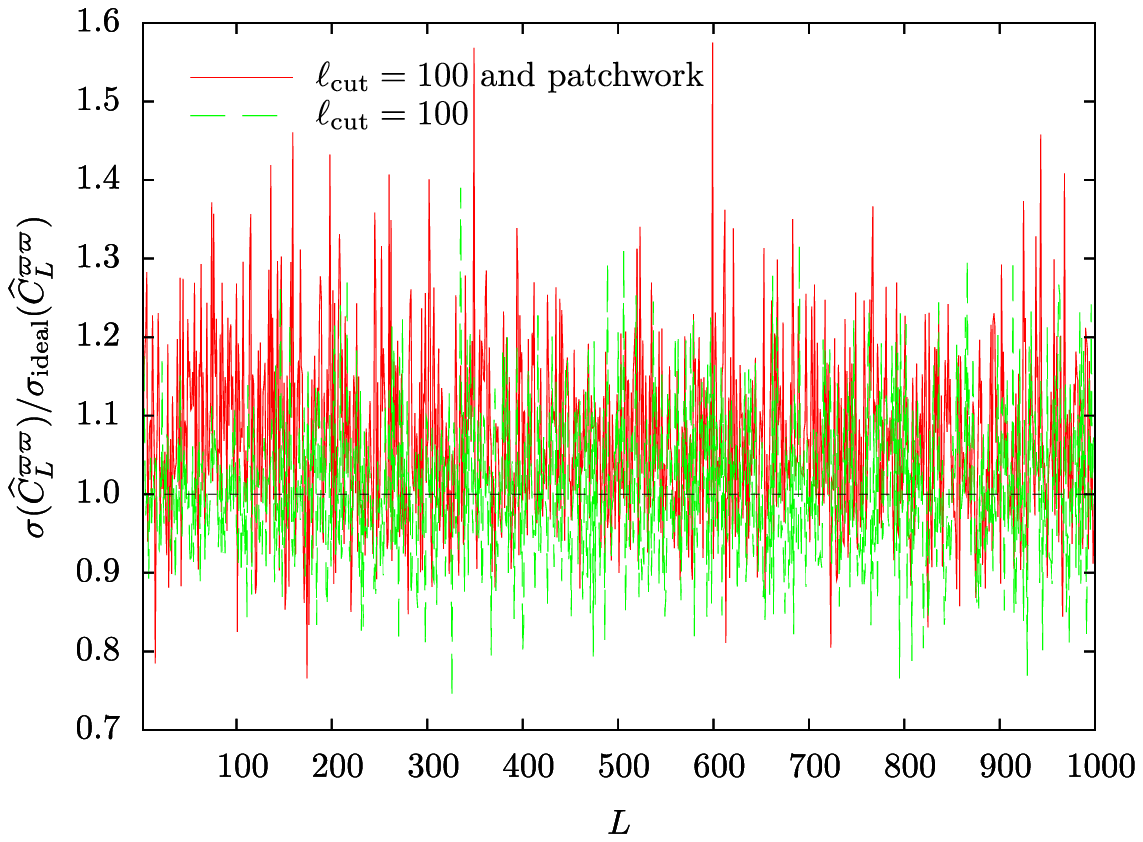}
\caption{
The variance of the gradient (left) and curl-mode (right) power spectrum estimators obtained 
from the patchwork maps ($\ell\rom{knee}=100$ and $\alpha=2$) with $\ell\rom{cut}=100$ (red) 
or that from coherent maps where $6\mu$K-arcmin noise is uniformly distributed 
but we apply $\ell\rom{cut}=100$ to the lensing reconstruction (green). 
For both cases, we divide the variance by that from the coherent maps without restriction 
of the Fourier modes in the lensing reconstruction. 
}
\label{Fig:Clxx:var}
\ec
\end{figure}

In Fig.~\ref{Fig:Clxx:var}, we show the variance of the power spectrum estimator for the gradient and curl mode obtained 
from the patchwork maps ($\ell\rom{knee}=100$ and $\alpha=2$) with $\ell\rom{cut}=100$  
or that from coherent maps where $6\mu$K-arcmin noise is uniformly distributed but we apply $\ell\rom{cut}=100$ to the lensing reconstruction. 
For both cases, we divide the variance by that from the coherent maps without restriction of the Fourier modes in the lensing reconstruction. 
In our case, there are two effects which affect the variance; the restriction of the Fourier 
modes in the lensing reconstruction, and the noisy reconstruction due to the leakage of large scale 
$1/f$ noise to small scales through the convolution of the subpatch window function (see Fig.~\ref{Fig:frac}). 
The results imply that the former effect is negligible if $\ell\rom{cut}=100$. 
On the other hand, for gradient mode, the variance from patchwork map is increased at smaller scales 
(i.e., the mean of the curve for the patchwork case tends to deviate from unity at smaller scales) 
where the cosmic variance of the lensing potential becomes negligible. 
Since we do not include curl mode signal, the variance for curl mode is affected by the latter effect at almost all scales. 
The increase of the variance is, however, only $\sim10\%$ even at noise dominated scales.

%% file: sec4.tex
\section{Cosmological applications of lensing observables from patchwork map} \label{sec.4}

In this section, we discuss cosmological applications of the reconstructed lensing potential from the patchwork of polarization maps. 
In the following analysis, we cross-correlate the reconstructed potential with (coherent) full-sky CMB temperature and polarization maps. 
We assume that the full-sky temperature map is obtained from the PLANCK experiment. Since our original motivation is to measure lensing signals in 
the CMB anisotropies economically, making the full-sky polarization map by a small-size satellite is complementary to our patchwork 
scheme. Taking account of this point, we assume that the full-sky polarization map is measured by LiteBIRD \footnote{\url{http://litebird.jp/}} which 
is a recently proposed small-size satellite observation with sensitivity high enough to measure the large scale lensing B-modes 
\footnote{PIXIE \cite{Kogut:2011xw} has similar experimental specification and is also suitable for our plan.}. 
For the simulations of such joint analysis, we additionally prepare $100$ realizations of the polarization (temperature) map in which 
the LiteBIRD (PLANCK) noise is generated as a Gaussian random field and then added to the polarization (temperature) map. 
The relevant noise level and beam size to LiteBIRD are $2\mu$K-arcmin and $30$ arcminutes FWHM, respectively \cite{LiteBIRD}, 
while the PLANCK noise level is computed according to Ref.~\cite{PLANCK}. 

\subsection{Temperature-lensing and E-mode-lensing cross-correlations} \label{sec:CrossCor}

\begin{figure} 
\bc
\includegraphics[width=75mm,clip]{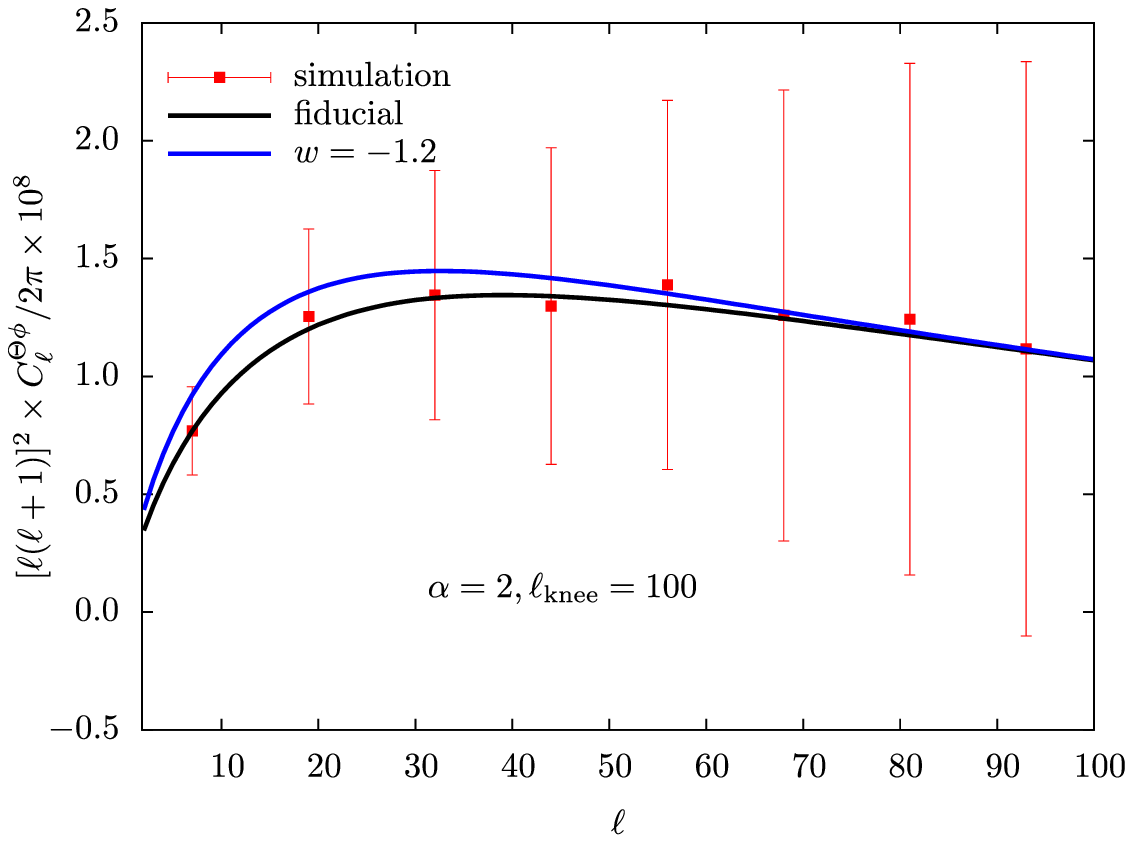}
\includegraphics[width=75mm,clip]{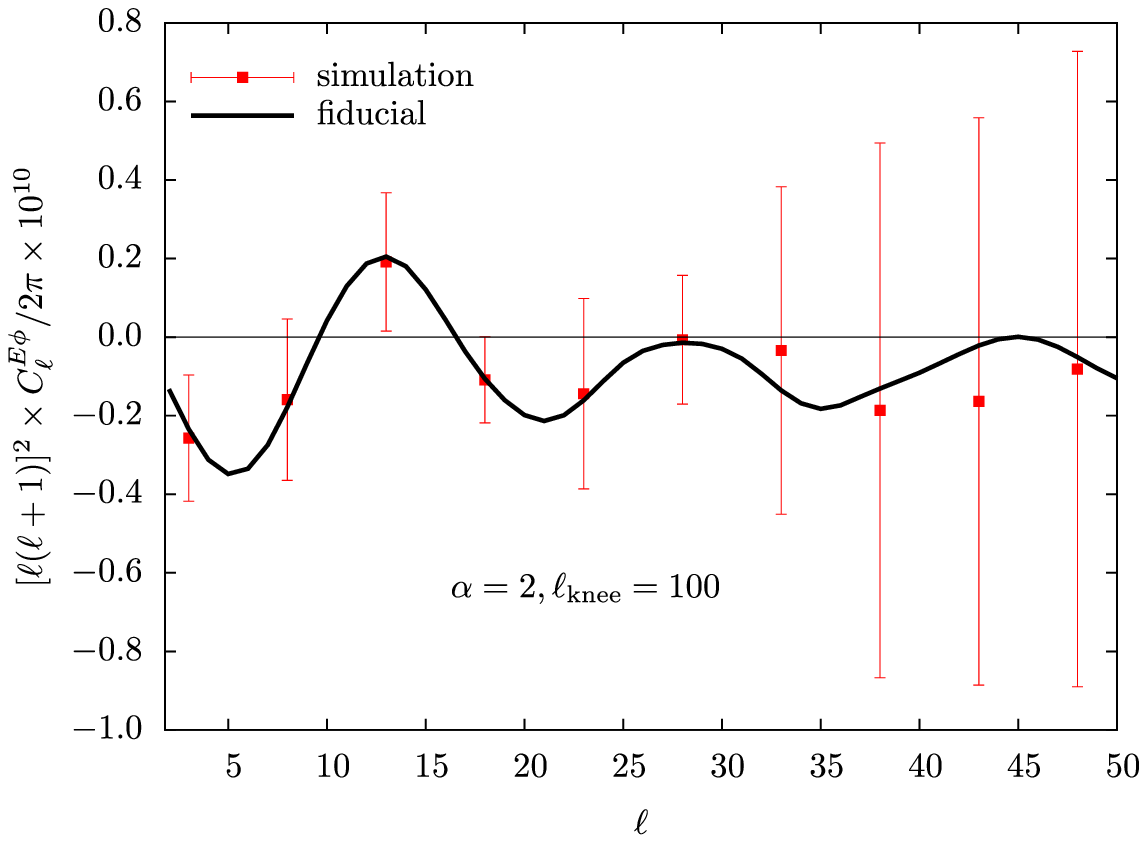}
\caption{
Temperature-lensing (left) and E-mode-lensing (right) cross-power spectrum from numerical simulations compared with the input power spectrum. 
In the simulations, PLANCK and LiteBIRD noises are added to the temperature and E-mode polarization, respectively. 
The noise level of the patchwork polarization map is set to $6\mu$K-arcmin. 
The error bar is computed from $100$ realizations of the numerical simulations. 
In the left panel, we also show the case with varying the dark-energy equation-of-state parameter $w=-1.2$ for comparison. 
}
\label{Fig:cross} 
\ec
\end{figure}

Since the reconstructed lensing potential is unbiased even at the largest scales, we can measure cross-power spectrum between the lensing potential and 
CMB temperature or E-mode polarization anisotropies which have large amplitudes on large scales ($\ell\lsim 100$). The galaxy-lensing cross-power spectrum 
would be also useful as a probe of the primordial non-Gaussianity. The gravitational potentials from the large-scale structure produces both the lensing effect 
and the temperature anisotropies, and the temperature-lensing cross-correlation would be a probe of the properties of the dark energy
\cite{Goldberg:1999xm,Seljak:1998nu}. The E-mode from the reionization also correlates with the deflection angle, producing E$\grad$ cross-power spectrum 
on large scales \cite{Lewis:2011fk}. To show the cross-power spectra, we first define the averaging factor of the power spectrum at each bin as 
\al{
	\widehat{A}^{X\grad}_b \equiv (\sigma^{X\grad}_b)^2\sum_{\ell=\ell^b\rom{min}}^{\ell^b\rom{max}} 
		\mS{B}^{X\grad}_{\ell} \frac{\hC^{X\grad}_{\ell}}{C^{X\grad}_{\ell}} 
	\,, \label{Eq:Ab} 
}
where the subscript $X$ is $\Theta/E$, the quantities $C^{X\grad}_{\ell}$ and $\hC^{X\grad}_{\ell}$ are the input and simulated cross-power spectrum, 
respectively. The multipoles $\ell^b\rom{min}$ and $\ell^b\rom{max}$ are the minimum and maximum multipole of the $b$-th bin, and the binning function 
and the variance of $\widehat{A}^{X\grad}_b$ at the $b$-th bin are given by 
\al{
	\mS{B}^{X\grad}_{\ell} = \frac{ ( 2\ell + 1 ) \, (C_{\ell}^{X\grad})^2 }
	{ (C^{\rm X\grad}_{\ell})^2 +(C^{\rm XX}_{\ell}+\mC{N}_{\ell})(C^{\grad\grad}_{\ell}+A_{\ell})} 
	\,, \qquad 
	\sigma^{X\grad}_b = \left\{\sum_{\ell} \mS{B}^{X\grad}_{\ell}\right\}^{-1/2} 
	\,, \label{Eq:bandpass-error} 
}
with $\mC{N}$ denoting the PLANCK (LiteBIRD) instrumental noise if $X=\Theta$ ($E$). Then we obtain the binned power spectrum at $b$-th bin as 
$\widehat{C}^{X\grad}_b=\widehat{A}^{X\grad}_bC^{X\grad}_{\ell\rom{b}}$ where $\ell\rom{b}=(\ell^b\rom{min}+\ell^b\rom{max})/2$. 

In Fig.~\ref{Fig:cross}, as a demonstration, we show the power spectra of temperature-lensing and E-mode-lensing cross-correlation. 
In computing the binned power spectrum, the multipole range between $\ell=2$ and $100$ is divided into $8$ bins for $\widehat{C}_b^{\Theta\grad}$ 
and $20$ bins for $\widehat{C}_b^{E\grad}$. The lensing potential is reconstructed using the procedure described in Sec.~\ref{sec.3}. 
The low-$\ell$ cut ($\ell\rom{cut}$) is set to $\ell\rom{knee}$ and the noise level of the patchwork map is $6\mu$K-arcmin. 
As expected, the cross-power spectra are consistent with the input power spectra. The signal-to-noise ratio of the cross-power spectrum is estimated as 
$\{\sum_b (\sigma_b^{X\grad})^{-2}\}^{1/2}$. Assuming the noise level of PLANCK, the signal-to-noise ratio of the temperature-lensing 
cross-correlation becomes $\sim 6.9\,\sigma$. On the other hand, by cross-correlating LiteBIRD polarization map with the reconstructed $\grad$ map, 
the E-mode and lensing cross-power spectrum would be detected with $\sim 2.4\,\sigma$ statistical significance. 

\subsection{Delensing B-mode polarization} \label{sec:delensing}

Once we obtain the lensing potential using the quadratic estimator, we can estimate the contribution of the lensing to the B-mode polarization based on 
Eq.~\eqref{Eq:Lensed-B}. This method is important to probe non-lensing B-modes and also used to estimate the lensing potential based on 
maximum-likelihood approach \cite{Hirata:2003ka}. In this paper, we follow the method described in Ref.~\cite{Smith:2010gu}: 
\al{
	\hB^{\rm lens}_{\ell m} &= -\iu \sum_{\ell'm'}\sum_{LM}
		\Wjm{\ell}{\ell'}{L}{m}{m'}{M}\mC{S}^{(-)}_{\ell\ell'L} \mC{W}^E_{\ell'}\mC{W}^{\grad}_L
		(\hE_{\ell'm'}\estg_{LM})^* 
	\,, \label{Eq:est-lensedB}
}
where we define the Wiener filter, $\mC{W}^{\grad}_L=C^{\grad\grad}_L/(C^{\grad\grad}_L+A_L)$ and 
$\mC{W}^{\rm E}_{\ell}=\CEE_{\ell}/\hCEE_{\ell}$ \cite{Seljak:2003pn,Smith:2010gu}. 
The residual B-mode polarization is then estimated from \cite{Smith:2010gu}
\al{
	\hB^{\rm res}_{\ell m} = \hB_{\ell m} - \hB^{\rm lens}_{\ell m}
	\,. \label{Eq:resB}
}

\subsubsection{Residual B-mode polarization} 

\begin{figure} 
\bc
\includegraphics[width=90mm,clip]{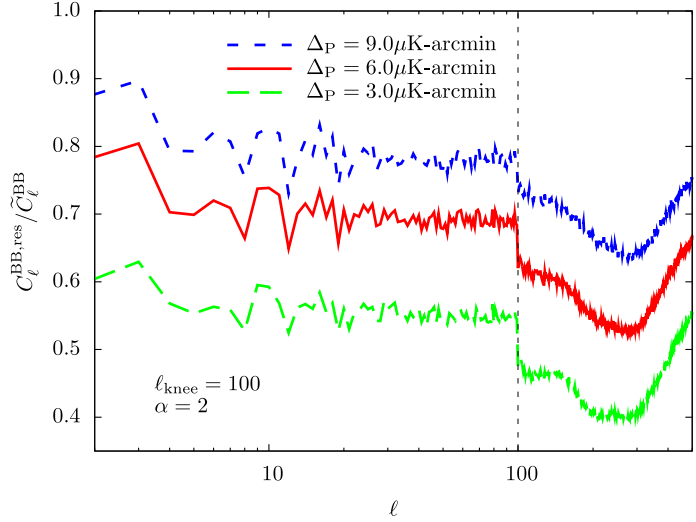}
\caption{
Residual fractions of the lensing B-modes. The residual fraction is defined as the ratio of the lensing B-mode spectrum after delensing to 
the original lensing B-mode spectrum. We show the cases with varying the polarization sensitivity $\Delta\rom{P}$. 
}
\label{Fig:delens} 
\ec
\end{figure}

In Fig.~\ref{Fig:delens}, to see feasibility of the above delensing algorithm, we show residual fraction of the lensing B-mode by delensing: 
$C_{\ell}^{\rm BB,res}/\tCBB_{\ell}$ (to clarify the delensing efficiency itself, $C_{\ell}$s in this factor do not include 
the noise power of LiteBIRD). By delensing, the lensing contributions in the B-mode power spectrum at $\ell< \ell\rom{knee}$ are subtracted by 
approximately $30\%$ in the case that the noise level of the patchwork map is $6\mu$K-arcmin. We checked that dependence on the parameters $\alpha$ 
and $\ell\rom{knee}$ is not so significant. This feature comes from the use of the EB-quadratic estimator in estimating the lensing potential 
(see appendix \ref{App:DB} for further explanations). The large discrepancy at $\ell\geq\ell\rom{knee}$ would be mainly reproduced by adding 
correction terms described in appendix \ref{App:DB} to Eq.~(2.12) of Ref.\cite{Smith:2010gu}. 

\begin{figure} 
\bc
\includegraphics[width=85mm,clip]{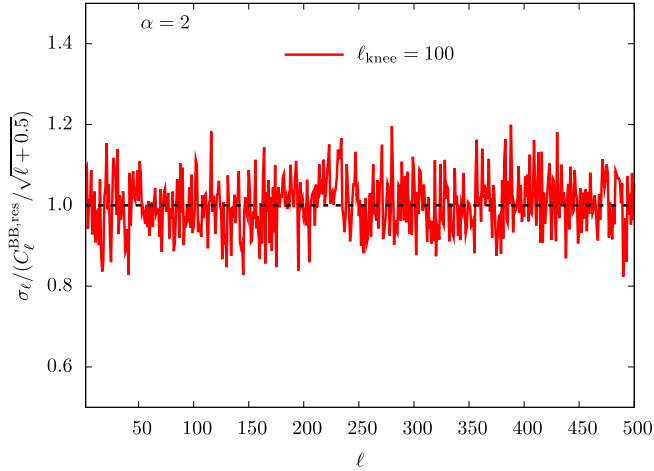}
\caption{
Variance of the residual B-mode power spectrum $\sigma_{\ell}$ obtained from $100$ realizations of the Monte Carlo simulation divided by 
$C_{\ell}^{\rm BB,res}/\sqrt{\ell+0.5}$, i.e., the variance obtained by assuming that the residual B-mode is Gaussian. Lensing reconstruction is 
performed from the patchwork maps with the noise level of $\Delta\rom{P}=6\mu$K-arcmin and the beam size of $4$ arcmin FWHM. 
}
\label{Fig:variance}
\ec
\end{figure}

In Fig.~\ref{Fig:variance}, we show variance of the residual B-mode power spectrum $\sigma_{\ell}$ obtained from $100$ realizations of 
the numerical simulations, divided by $C_{\ell}^{\rm BB, res}/\sqrt{\ell+0.5}$, i.e., the variance obtained by assuming that the residual B-mode is Gaussian. 
To evaluate the Gaussian variance $C_{\ell}^{\rm BB,res}/\sqrt{\ell+0.5}$, we use the residual B-mode power spectrum from the numerical simulations. 
The variance of the angular power spectrum does not significantly deviate from the Gaussian variance. 

\subsubsection{Expected constraints on primordial gravitational waves with LiteBIRD} 

\begin{table}
\bc
\caption{
Expected $1\,\sigma$ constraints on $r$ of LiteBIRD observation without delensing (no delensing) and with delensing based on the joint analysis 
with ground-based experiments of $\Delta\rom{P} = 6\mu$K-arcmin or $3\mu$K-arcmin. We also show the case in which $90$\% of 
the lensing components are removed ($90$\% delensing) and an idealistic case where the lensing component is completely removed (LiteBIRD limit). 
In all cases, we use the B-mode multipoles up to $\ell=300$ where the LiteBIRD noise becomes dominant. The tensor tilt is fixed to $n\rom{t}=-r/8$. 
}
\label{Table:r} \vs{0.5}
\begin{tabular}{c|ccc|cc} \hline 
$r$ & {\small no delensing} & {\small $6\mu$K-arcmin} & {\small $3\mu$K-arcmin} & {\small $90$\% delensing} & {\small LiteBIRD limit} \\ \hline 
$0.2$ & $2.3\times 10^{-3}$ & $2.2\times 10^{-3}$ & $2.1\times 10^{-3}$ & $1.7\times 10^{-3}$ & $1.6\times 10^{-3}$ \\ 
$0.02$ & $4.8\times 10^{-4}$ & $4.1\times 10^{-4}$ & $3.8\times 10^{-4}$ & $2.7\times 10^{-4}$ & $2.4\times 10^{-4}$ \\ 
$0.002$ & $2.3\times 10^{-4}$ & $1.9\times 10^{-4}$ & $1.6\times 10^{-4}$ & $8.7\times 10^{-5}$ & $6.8\times 10^{-5}$ \\ \hline
\end{tabular}
\ec
\end{table}

\begin{table}
\bc
\caption{
Same as Table \ref{Table:r}, but for marginalizing the tensor tilt $n\rom{t}$. 
Note that the fiducial values are $r=0.2$ and $n\rom{t}=-r/8=-0.025$. 
}
\label{Table:r-nt} \vs{0.5}
\begin{tabular}{c|ccc|cc} \hline 
& {\small no delensing} & {\small $6\mu$K-arcmin} & {\small $3\mu$K-arcmin} & {\small $90$\% delensing} & {\small LiteBIRD limit} \\ \hline 
$\sigma(r)$ & 0.0095 & 0.0087 & 0.0084 & 0.0072 & 0.0068 \\ 
$\sigma(n\rom{t})$ & 0.028 & 0.025 & 0.024 & 0.019 & 0.018 \\ \hline 
\end{tabular}
\ec
\end{table}

As an example of non-lensing B-mode probes, let us discuss the expected constraints on the amplitude and shape of the spectrum of primordial 
gravitational waves using a joint delensing analysis of LiteBIRD and ground-based CMB experiments. The polarization map ``to be delensed'' is provided 
by LiteBIRD and the lensing pontential map is reconstructed from the patchrork polarization map measured by ground-based experiments. 
Since satellite missions usually have the advantage of relatively low noise level, such joint analysis would make good synergy. 

We characterize the primordial tensor power spectrum as $P_t(k)=rA\rom{s}(k/k_0)^{n\rom{t}}$ where $r$ is the tensor-to-scalar ratio and $n\rom{t}$ 
is the tensor spectral index. $A\rom{s}$ is the amplitude of the primordial scalar power spectrum at $k_0$. 
Here, we assume that these two parameters satisfy the consistency relation $n\rom{t}=-r/8$. The pivot scale $k_0$ is set to $k=0.05$ Mpc$^{-1}$. 

Before moving to discuss the expected constraints on $r$ and $n\rom{t}$, we comment on the partial subtraction of 
the primary B-mode \cite{Hirata:2003ka,Teng:2011xc}. The origin of the primary B-mode subtraction is the same as that of the step feature in the 
residual B-mode shown in the previous section (see appendix \ref{App:DB}). This subtraction would also lead to a non-trivial likelihood for the residual B-mode. 
For the primordial gravitational waves, however, as shown in Ref.~\cite{Seljak:2003pn}, this partial subtraction can be simply evaded if we incorporate only 
the B-mode multipoles at $\ell\geq\ell\rom{cut}$ into the lensing reconstruction analysis and delense the multipoes at $\ell<\ell\rom{cut}$. 
Information of the lensing potential mostly comes from structure of arcminute scales and LiteBIRD's beam size is $30$ arcminutes FWHM. 
In the analysis below, we assume $\ell\rom{cut}=300$. This choice of $\ell\rom{cut}$ simultaneously mitigates the bias in the lensing potential due to 
reconstruction from the patchwork map (see Sec.~\ref{sec:rec}). 

The expected constraints on $r$ and $n\rom{t}$ are computed based on the Fisher matrix defined as 
\al{
	F_{ij} \equiv \sum_{\ell=2}^{\ell\rom{cut}-1} 
		\frac{2\ell+1}{2[C_{\ell}^{\rm BB,res}]^2}
		\PD{C^{\rm BB,res}_{\ell}}{p_i}\PD{C^{\rm BB,res}_{\ell}}{p_j}
	\,, 
}
where $p_i=r$ or $n\rom{t}$. For simplicity, we ignore the contributions of the Galactic foreground emission 
\footnote{Note that, based on Ref.~\cite{Katayama:2011eh}, the foreground components can be suppressed at few percent.}. 
The derivatives are evaluated by finite difference method. The residual B-mode power spectrum $C^{\rm BB,res}_{\ell}$ is obtained from 
our numerical simulation. The fiducial values are $r=0.2$ and $n\rom{t}=-r/8=0.025$. The expected $1\,\sigma$ error on a parameter $p_i$ is 
computed as $\sigma(p_i)=\sqrt{\{F^{-1}\}_{ii}}$. 

Although the recent BICEP2 results show $r=0.2^{+0.07}_{-0.05}$ \cite{Ade:2014xna}, the WMAP and PLANCK constraint on $r$ is 
$r\lsim 0.1$ \cite{Ade:2013zuv}. There are many on-going and future experiments to show whether this discrepancy is real or not. 
For this reason, let us first consider the expected constraints for several values of $r$ with the tensor tilt fixed as $n\rom{t}=-r/8$. 
The resultant constraints are shown in Table \ref{Table:r}. We show the cases without delensing (no delensing), and 
with delensing based on the joint analysis with ground-based experiments of $\Delta\rom{P} = 6\mu$K-arcmin or $3\mu$K-arcmin. 
We also show the case if the lensing B-mode is removed by $90$\% or perfectly removed (LiteBIRD limit). 

Next we consider the case if the tensor-to-scalar ratio is confirmed as $r\sim 0.2$. In Table \ref{Table:r-nt}, we give the expected constraints on $r$ 
and $n\rom{t}$. Without delensing, the expected error of the tensor tilt $\sigma(n\rom{t})$ is $0.028$. Delensing using a $6\mu$K-arcmin experiment 
would improve the constraints on $n\rom{t}$ by $8$\%. If the delensing effifiency of $90$\% is attainable, the expected error reduces 
to $\sigma(n\rom{t})=0.019$. 

%% file: sec5.tex
\section{Summary and Discussion} \label{sec.5} 

We have explored for the first time the feasibility of lensing reconstruction from the patchwork 
of CMB polarization maps. 
The B-mode power spectrum of the patchwork map contains the residual bias and its significance 
depends on the subpatch size but slightly on the shape parameter $\alpha$. 
On the other hand, the bias on the estimated lensing potential would be negligible 
if the B-modes at $\ell<\ell\rom{knee}$ are removed in the lensing reconstruction. 
We also performed a null consistency test of curl mode and found that 
we must care about the N1-bias on curl mode as in the case of 
temperature-based reconstruction \cite{vanEngelen:2012va,BenoitLevy:2013bc}. 
Based on these analyses, we discussed cosmological applications of the reconstructed 
lensing potential from the patchwork map. 
Since the lensing potential is unbiased even at the largest scale, we would measure, 
for example, the temperature-lensing and E-mode-lensing cross-power spectrum. 
Delensing of the lensing B-mode was also considered. 
We investigated the efficiency of delensing and found that the reconstructed potential could be 
used for restoring the lensing B-modes. 
The variance of the residual B-mode did not significantly deviate from the Gaussian variance 
($C_{\ell}^{\rm BB,res}/\sqrt{\ell+0.5}$). 
Our parameter forecast based on Fisher matrix analysis showed that the expected 
LiteBIRD constraint on $r$ was improved especially in the cases of small $r$. 
Specifically, the expected error of $r$ reduces by $17$\% ($30$\%) in the case of $r=0.002$ 
if we delense the polarization map measured by LiteBIRD using the lensing potential 
reconstructed from the patchwork map with the noise level of $6\mu$K-arcmin ($3\mu$K-arcmin). 
We also estimated the LiteBIRD constraint on $n\rom{t}$ in the case of $r=0.2$ which was 
the value claimed by BICEP2. 
The constraint is improved by $8$\% ($12$\%) if we assume the patchwork map with 
the noise level of $6\mu$K-arcmin ($3\mu$K-arcmin). 

We comment on the case if the subpatch size becomes larger 
than the case considered in this paper. In this case, we can use B-modes on larger scales 
for unbiased lensing reconstruction. 
Besides, the leakage of 1/f noise to small scales decreases, 
which would reduce the variance of the power spectrum estimator. 
For delensing LiteBIRD B-mode, on the other hand, such large subpatch 
size would not improve the delensing efficiency 
if we mitigate the delensing bias by filtering out of large scale
B-modes.

In our study, to focus on the effect of the baseline uncertainty on the lensing reconstruction, 
we used some simplifications and assumptions on the patchwork map. 
For example, we assumed that the map of each subpatch was measured through isotropic sky scanning. 
In actual situations, anisotropic sky scanning make anisotropic deficits in Fourier space. 
There are also other practical issues associated with this reconstruction procedure such as offsets 
of subpatch locations and mismatches in relative gain between subpatches. 
We assumed our patchwork map covered the whole sky. 
The finite survey area and point source mask would also be sources of systematic errors 
in lensing reconstruction. 
In our analysis, we assumed experiments which were originally designed
to make a patchwork map of CMB polarization. If we consider more general
case where a patchwork map contains subpatches observed by independent experiments with different
experimental specifications, there would be several possible systematics such as
different beam, noise levels and window functions charactering each subpatch. 
We checked that the ``variance'' of the residual B-mode power spectrum 
did not so deviate from that obtained assuming the residual B-modes were Gaussian. 
The likelihood of the residual B-mode multipoles, however, have not been thoroughly explored. 
Further investigation of the above issues will be presented in our future work.

%% file: appA.tex
\section{Delensing bias} \label{App:DB}

In this section, we explain the step feature in the residual B-mode power spectrum also discussed in Ref.~\cite{Teng:2011xc} as delensing bias. 
In the following calculations, we discuss expression for the residual B-mode power spectrum. Note that we frequently use the orthogonality relation \cite{QTAM} 
\al{
	\sum_{m_1m_2}\Wjm{\ell_1}{\ell_2}{\ell_3}{m_1}{m_2}{m_3} \Wjm{\ell_1}{\ell_2}{\ell'_3}{m_1}{m_2}{m_3'} 
		= \frac{\delta_{\ell_3\ell_3'}\delta_{m_3m_3'}}{2\ell_3+1} 
	\,, 
}
and the symmetric property of the Wigner 3-j symbols \cite{QTAM}: 
\al{
	\Wjm{\ell_1}{\ell_2}{\ell_3}{m_1}{m_2}{m_3} &= \Wjm{\ell_2}{\ell_3}{\ell_1}{m_2}{m_3}{m_1} \,, 
	\notag \\ 
	\Wjm{\ell_1}{\ell_2}{\ell_3}{m_1}{m_2}{m_3} &= (-1)^{\ell_1+\ell_2+\ell_3}\Wjm{\ell_2}{\ell_1}{\ell_3}{m_2}{m_1}{m_3} 
	\,. 
}

\subsection{Lensing B-mode estimator} 

Let us consider if we use the EB-estimator for the lensing reconstruction: 
\al{
	[\estg^{\rm EB}_{LM}]^* 
		&= A^{\rm EB}_L\sum_{\ell\ell'mm'}\Wjm{\ell}{\ell'}{L}{m}{m'}{M}(g^{\rm EB}_{\ell'\ell L})^*\hB_{\ell m}\hE_{\ell'm'} 
	\,, \label{Eq:estg-EB} 
}
where $g^{\rm EB}_{\ell'\ell L}=-\iu\mC{S}^{(-)}_{\ell\ell'L}\mC{W}^E_{\ell'}/\hCBB_{\ell}$. 
Using the expression of the EB-estimator \eqref{Eq:estg-EB}, the lensing B-mode estimator becomes 
\al{
	\hB^{\rm lens}_{\ell m} &= \sum_{L\ell'}\mC{W}^{\grad}_L A^{\rm EB}_L 
			\frac{(f^{\rm EB}_{\ell'\ell L})^*}{\hC^{\rm EE,LB}_{\ell'}}\sum_{Mm'}\Wjm{\ell}{\ell'}{L}{m}{m'}{M}
	\notag \\ 
		&\qquad \times 
			\sum_{\ell_1\ell'_1}\sum_{m_1m'_1}\Wjm{\ell_1}{\ell'_1}{L}{m_1}{m'_1}{M}
			(g^{\rm EB}_{\ell'_1\ell_1L})^*(\hE^{\rm LB}_{\ell'm'})^*\hB_{\ell_1m_1}\hE_{\ell'_1m'_1} 
	\,, \label{Eq:lensing-B}
}
where $\hE^{\rm LB}_{\ell m}$ is the E-mode of LiteBIRD observation, and $\hC^{\rm EE,LB}_{\ell}$ is the LiteBIRD E-mode angular power spectrum. 
We first consider the following term involved in the bispectrum, which is obtained by taking ensemble average over E-mode polarization: 
\al{
	(\hE^{\rm LB}_{\ell'm'})^*\hB_{\ell_1m_1}\hE_{\ell'_1m'_1} &\ni \delta_{\ell'\ell'_1}\delta_{m'm'_1}\tCEE_{\ell'}\hB_{\ell_1m_1} 
	\,. \label{Eq:decompose}
}
Substituting the first term in the r.h.s. of the above relation into Eq.~\eqref{Eq:lensing-B}, we obtain 
\al{
	\hB^{\rm lens}_{\ell m} &\ni \sum_{\ell_1m_1}\sum_{L\ell'}\mC{W}^{\rm E,LB}_{\ell'}\mC{W}_L^{\grad}A^{\rm EB}_L
			\sum_{Mm'}\Wjm{\ell}{\ell'}{L}{m}{m'}{M} \Wjm{\ell_1}{\ell'}{L}{m_1}{m'}{M} 
			(f^{\rm EB}_{\ell'\ell L}g^{\rm EB}_{\ell'\ell L})^* \hB_{\ell_1 m_1}
	\notag \\ 
		&= \hB_{\ell m} \sum_{\ell_1m_1}\frac{\delta_{\ell\ell_1}\delta_{mm_1}}{2\ell+1}
			\sum_{L\ell'}\mC{W}^{\rm E,LB}_{\ell'} \mC{W}_L^{\grad} A^{\rm EB}_L (f^{\rm EB}_{\ell'\ell L}g^{\rm EB}_{\ell'\ell L})^* 
	\notag \\ 
		&= \hB_{\ell m} \frac{1}{\hC_{\ell}^{\rm BB}}
			\sum_{\ell_1m_1}\frac{\delta_{\ell\ell_1}\delta_{mm_1}}{2\ell+1}\sum_{L\ell'}(\mC{S}^{(-)}_{\ell\ell'L})^2 
			[\mC{W}^{\rm E,LB}_{\ell'}\mC{W}^{\rm E}_{\ell'}\CEE_{\ell'}] [\mC{W}_L^{\grad}A^{\rm EB}_L]
	\notag \\ 
		&\equiv \mC{D}_{\ell} \hB_{\ell m} 
	\,, \label{Eq:lensing-B:d}
}
where we define $\mC{W}^{\rm E,LB}_{\ell}=\CEE_{\ell}/\hC^{\rm EE,LB}_{\ell}$ and 
\al{
	\mC{D}_{\ell} \equiv \frac{1}{\hC_{\ell}^{\rm BB}}
		\sum_{\ell_1m_1}\frac{\delta_{\ell\ell_1}\delta_{mm_1}}{2\ell+1}\sum_{L\ell'}(\mC{S}^{(-)}_{\ell\ell'L})^2 
		[\mC{W}^{\rm E,LB}_{\ell'}\mC{W}^{\rm E}_{\ell'}\CEE_{\ell'}] [\mC{W}_L^{\grad}A^{\rm EB}_L] 
	\,. 
}
Note that, if we do not use the B-modes at $\ell<\ell\rom{min}$ in the lensing reconstruction, the quantity \eqref{Eq:lensing-B:d} becomes zero 
at $\ell<\ell\rom{min}$. This leads to the discontinuity in the residual B-mode spectrum. In actual situations, the primary B-mode is involved 
in $\hB_{\ell m}$, which causes the partial subtraction of primary B-mode signal as pointed out in Ref.~\cite{Teng:2011xc}. 

On the other hand, a term in which the correlation between $E^{\rm LB}$ and $\estg$ is ignored is given by 
\al{
	\hB^{\rm lens}_{\ell m} 
		&\ni \sum_{L\ell'}\mC{W}^{\grad}_L \frac{(f^{\rm EB}_{\ell'\ell L})^*}{\hC^{\rm EE,LB}_{\ell'}}
			\sum_{Mm'} \Wjm{\ell}{\ell'}{L}{m}{m'}{M}[(\hE^{\rm LB}_{\ell'm'}\estg_{LM}^{\rm EB})^*]\rom{d} 
		\equiv \tB^{\rm W}_{\ell m}
	\,. 
}
Combining the above equation and Eq.~\eqref{Eq:lensing-B:d}, we obtain 
\al{
	\hB^{\rm lens}_{\ell m} \simeq \tB^{\rm W}_{\ell m} + \mC{D}_{\ell} \hB_{\ell m}
	\,. \label{Eq:lensing-B:tot}
}
As shown below, the second term is required to explain the step feature in the residual B-mode power spectrum. Hereafter, we call the second term 
$\mC{D}_{\ell}\hB_{\ell m}$ as delensing bias. Note that, in the above expression, the delensing bias is vanished if we estimate the lensing potential from 
other experiments, e.g., surveys of cosmic infrared background, galaxy weak lensing and so on. 

\subsection{Residual B-mode power spectrum} 

Now we consider the angular power spectrum of the residual B-mode in the presence of the delensing bias: 
\al{
	\ave{|\hB^{\rm res}_{\ell m}|^2} 
		&= \ave{|\hB^{\rm LB}_{\ell m} - \tB^{\rm W}_{\ell m} - \mC{D}_{\ell}\hB_{\ell m}|^2} 
	\notag \\ 
		&= \ave{|\hB^{\rm LB}_{\ell m} - \tB^{\rm W}_{\ell m}|^2} - 2\mC{D}_{\ell}\ave{|\hB^{\rm LB}_{\ell m}(\hB_{\ell m})^*|} 
		+ 2\mC{D}_{\ell}\ave{|\hB^{\rm W}_{\ell m}(\hB_{\ell m})^*|} + \mC{D}_{\ell}^2\ave{|\hB_{\ell m}|^2} 
	\,. \label{Eq:lensing-B:auto}
}
Here we denote the B-mode polarization observed by LiteBIRD as $\hB^{\rm LB}_{\ell m}$. As shown in the above equation, the delensing bias introduces 
the second, third and forth terms. Note that the delensing bias is proportional to the B-mode polarization, and correlates with the B-mode to be delensed, 
i.e., LiteBIRD B-mode. In other words, the delensing bias leads to a realization-dependent subtraction of the B-mode to be delensed. Denoting 
\al{
	\tC_{\ell}^{\rm BB,W} \equiv \ave{|\tB^{\rm W}_{\ell m}|^2} 
		= \frac{1}{2\ell+1}\sum_{\ell' L} (\mC{S}^{(-)}_{\ell\ell'L})^2 \mC{W}^{\rm E,LB}_{\ell'}\mC{W}_L^{\grad}\CEE_{\ell'}C_L^{\grad\grad}
	\,, 
}
the first term is given by \cite{Smith:2010gu} 
\al{
	\ave{|\hB^{\rm LB}_{\ell m}-\tB^{\rm W}_{\ell m}|^2} 
		&= \ave{|\hB^{\rm LB}_{\ell m}|^2} -2\ave{\hB^{\rm LB}_{\ell m}(\tB^{\rm W}_{\ell m})^*} + \ave{|\tB^{\rm W}_{\ell m}|^2} 
	\notag \\ 
		&= \hC_{\ell}^{\rm BB,LB} - 2\tC_{\ell}^{\rm BB,W} + \tC_{\ell}^{\rm BB,W}
		= \hC_{\ell}^{\rm BB,LB} - \tC_{\ell}^{\rm BB,W} 
	\,. \label{Eq:rClBB:Smith12}
}
The other terms due to the presence of the delensing bias become 
\al{
	\ave{|\hB^{\rm LB}_{\ell m}\hB^*_{\ell m}|} &= \tCBB_{\ell}
	\,, \\ 
	\ave{|\tB^{\rm W}_{\ell m}\hB^*_{\ell m}|} &= \tC_{\ell}^{\rm BB,W}
	\,, \\ 
	\ave{|\hB_{\ell m}|^2|} &= \hC_{\ell}^{\rm BB}
	\,. 
}
By combining the above equations, we obtain the expression for the residual B-mode power spectrum in the presence of the delensing bias as 
\al{
	C^{\rm BB,res}_{\ell} &= \hC^{\rm BB,LB}_{\ell} - \tC_{\ell}^{\rm BB,W} - 2\mC{D}_{\ell}\tCBB_{\ell} 
		+ 2\mC{D}_{\ell}\tC_{\ell}^{\rm BB,W} + \mC{D}_{\ell}^2 \hC_{\ell}^{\rm BB} 
	\,. \label{Eq:rClBB:Theo}
}

\begin{figure} 
\bc
\includegraphics[width=100mm,clip]{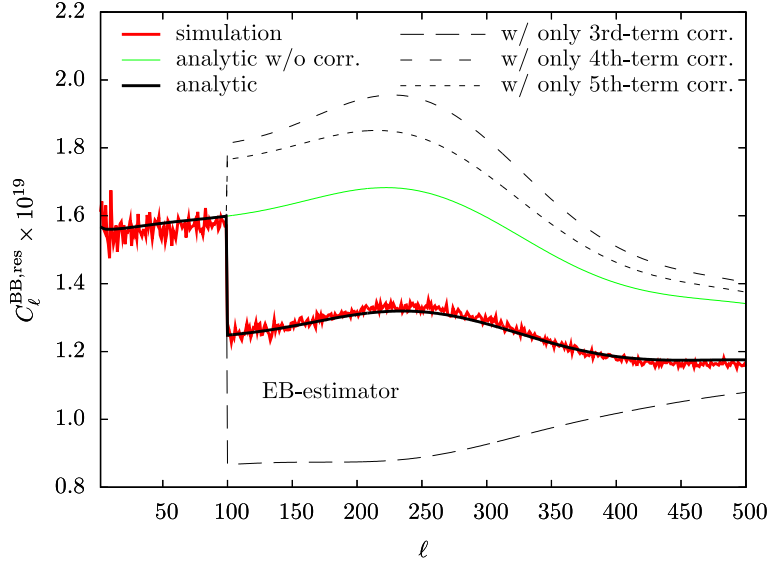}
\caption{
The residual B-mode power spectrum obtained from $100$ realizations of numerical simulation (red solid) compared with analytic expressions given in 
Ref.~\cite{Smith:2010gu} (analytic w/o corr.) and Eq.~\eqref{Eq:rClBB:Theo} (analytic). Note that, to match the setup applied to Fig.\ref{Fig:delens}, 
we assume $r=0$ and the LiteBIRD instrumental noise in B-mode is removed. To show significance of the correction terms, we also show power spectra 
using only $2\mC{D}_{\ell}\tCBB_{\ell}$ (w/ only 3rd-term corr.), $2\mC{D}_{\ell}\tC_{\ell}^{\rm BB,W}$ (w/ only 4th-term corr.) 
or $\mC{D}_{\ell}^2 \hC_{\ell}^{\rm BB}$ (w/ only 5th-term corr.) as a correction of the delensing bias term in Eq.~\eqref{Eq:rClBB:Theo}. 
In the numerical simulation, the lensing potential is reconstructed with the EB-quadratic estimator, and we add white noise as a random Gaussian field 
assuming our fiducial noise level and beam size ($\Delta\rom{P}=6\mu$K-arcmin and $\theta=4$ arcmin). 
}
\label{Fig:rClBB_EB}
\ec
\end{figure}

\begin{figure} 
\bc
\includegraphics[width=100mm,clip]{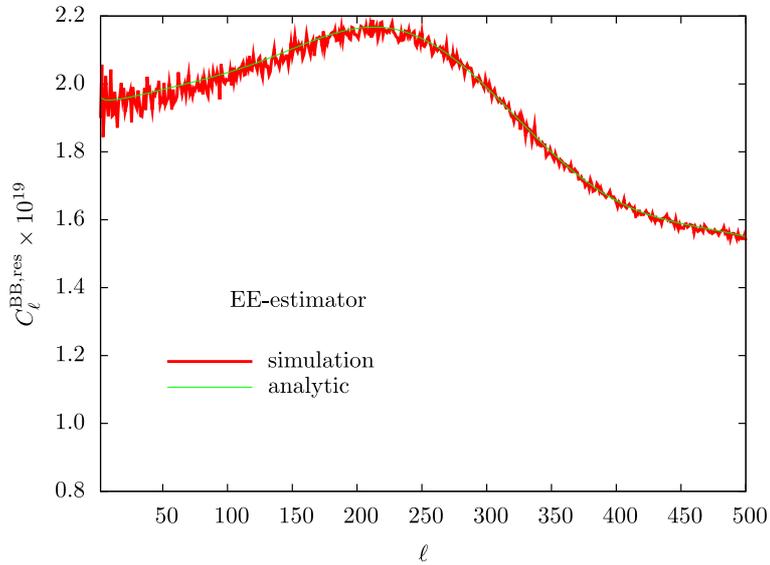}
\caption{
Same as Fig.~\ref{Fig:rClBB_EB} but for the case with the EE-quadratic estimator. 
}
\label{Fig:rClBB_EE}
\ec
\end{figure}

In Fig.~\ref{Fig:rClBB_EB}, we compare the theoretical power spectrum \eqref{Eq:rClBB:Theo} with the residual B-mode from the numerical simulation with 
a white noise generated as a random Gaussian field assuming the noise level of $\Delta\rom{P}=6\mu$K-arcmin and the beam size of $4$ arcminutes FWHM. 
To match the setup applied to Fig.~\ref{Fig:delens}, we assume $r=0$ and the LiteBIRD instrumental noise in B-mode is removed. As shown in 
Fig.~\ref{Fig:rClBB_EB}, the resultant residual B-mode power spectrum is suppressed at $\ell\geq\ell\rom{cut}$. This implies that the third term 
in Eq.~\eqref{Eq:rClBB:Theo} which comes from the correlations between the delensing bias and LiteBIRD B-mode is significant compared to other bias terms. 
On the other hand, in Fig.~\ref{Fig:rClBB_EE}, we show the case with the EE quadratic estimator for the lensing reconstruction. For EE-estimator, 
since $\mC{D}_{\ell}=0$, the analytic power spectrum of Eq.~\eqref{Eq:rClBB:Smith12} is in good agreement with the results of the numerical simulations.